\documentclass[useAMS,usenatbib,usegraphicx]{mn2e}
\usepackage{amssymb}
\usepackage{amsmath}

\title[Hunting for dark halo substructure]{Hunting for dark halo substructure using submilliarcsecond-scale observations of macrolensed radio jets}
\author[Zackrisson et al.]{Erik Zackrisson$^{1}$\thanks{E-mail: ez@astro.su.se}, Saghar Asadi$^{1}$, Kaj Wiik$^3$, Jakob J\"onsson$^2$, Pat Scott$^4$,\newauthor Kanan K. Datta$^1$, Martina M. Friedrich$^1$, Hannes Jensen$^1$, Joel Johansson$^2$,\newauthor Claes-Erik Rydberg$^1$, Andreas Sandberg$^1$\\  
$^1$Department of Astronomy, Stockholm University, Oscar Klein Center, AlbaNova, Stockholm SE-106 91, Sweden\\
$^2$Department of Physics, Stockholm University, Oscar Klein Center, AlbaNova, Stockholm SE-106 91, Sweden\\
$^3$Tuorla Observatory, Department of Physics and Astronomy, University of Turku, V\"ais\"al\"antie 20, FI-215 00 Piikki\"o, Finland\\
$^4$Department of Physics, McGill University, 3600 Rue University, Montr\'eal, QC H3A 2T8, Canada}

\begin{document}

\date{Accepted ... Received ...; in original form ...}

\pagerange{\pageref{firstpage}--\pageref{lastpage}} \pubyear{2010}

\maketitle

\label{firstpage}

\begin{abstract}
Dark halo substructure may reveal itself through secondary, small-scale gravitational lensing effects on light sources that are macrolensed by a foreground galaxy. Here, we explore the prospects of using Very Long Baseline Interferometry (VLBI) observations of multiply-imaged quasar jets to search for submilliarcsecond-scale image distortions produced by various forms of dark substructures in the $10^3$--$10^{8}\ M_\odot$ mass range. We present lensing simulations relevant for the angular resolutions attainable with the existing European VLBI Network (EVN), the global VLBI array, and an upcoming observing mode in which the Atacama Large Millimeter Array (ALMA) is connected to the global VLBI array. While observations of this type would not be sensitive to standard cold dark matter subhalos, they can be used to detect the more compact forms of halo substructure predicted in alternative structure formation scenarios. By mapping $\approx 5$ strongly lensed systems, it should be possible to detect or robustly rule out primordial black holes in the $10^3$--$10^6\ M_\odot$ mass range if they constitute $\gtrsim 1\%$ of the dark matter in these lenses. Ultracompact minihalos are harder to detect using this technique, but $10^6$--$10^8\ M_\odot$ ultracompact minihalos could in principle be detected if they constitute $\gtrsim 10\%$ of the dark matter.
\end{abstract}


\section{Introduction}
\label{intro}
A generic prediction of the standard cold dark matter (CDM) scenario is that a substantial fraction of the total mass of galaxy-sized dark matter halos \citep[$\sim 10\%$; ][]{Gao et al.,Maciejewski et al.} should be in the form of bound substructures (a.k.a. subhalos or subclumps) left over from the process of hierarchical assembly. The fact that the number of substructures seen in CDM simulations greatly outnumber the satellite galaxies detected in the vicinity of the Milky Way and Andromeda constitutes the so-called ``missing satellite problem'' \citep{Klypin et al.,Moore et al.}. While it has been argued that astrophysical processes that quench star formation in low-mass halos may explain this discrepancy \citep[e.g.][]{Maccio et al. b,Font et al.}, this implies that a vast population of extremely faint or completely dark substructures should be awaiting discovery in the halos of galaxies. Provided that CDM is in the form of Weakly Interacting Massive Particles (WIMPs), these subhalos are in principle detectable by the Fermi Gamma-ray Space Telescope because of their annihilation fluxes. However, Fermi has so far failed to detect any unambigious signal from such objects \citep[e.g.][ but see \citealt{Bringmann et al. b} and \citealt{Su & Finkbeiner} for a different view]{Belikov et al.,Zechlin et al.,Hooper & Linden}.

Gravitational lensing may provide an independent test for the presence of dark halo substructures \citep[for a review, see][]{Zackrisson & Riehm}. A foreground galaxy that happens to be aligned with a background light source can produce multiple images of the background object, with a typical image separation of $\sim 1\arcsec$ (an effect known as strong lensing or macrolensing). While simple, smooth models of galaxy lenses are usually able to reproduce the positions of these macroimages, their observed flux ratios are more difficult to explain. Such flux-ratio violations have been interpreted as evidence of substantial small-scale structure within the main lens
\citep[e.g.][]{Mao & Schneider,Chiba,Keeton et al.,Kochanek & Dalal}. A notable problem with this picture is that current CDM simulations predict too little substructure to explain many of these flux-ratio violations (e.g. \citealt{Maccio & Miranda,Xu et al. a,Chen et al.} -- but see \citealt{Metcalf & Amara}), possibly pointing to a considerable contribution from low-mass halos elsewhere along the line of sight \citep{Xu et al. b} or some additional form of substructure (dark or luminous) within the lens. 

A slightly different lensing approach exploits the small-scale distortions that halo substructure is expected to introduce in the morphologies of extended macroimages.  Substructures of mass $\gtrsim 10^8\ M_\odot$ can perturb gravitational arcs and Einstein rings on scales resolvable with the {\it Hubble Space Telescope} \citep[][]{Vegetti & Koopmans a,Vegetti & Koopmans b} and  detections of $\sim 10^9$--$10^{10}\ M_\odot$ objects have already been made this way \citep{Vegetti et al. a,Vegetti et al. b,Vegetti et al. c}. In line with the flux ratio anomaly results, these observations seem to suggest a subhalo mass fraction that is significantly higher than predicted by standard CDM, and possibly also a flatter subhalo mass function slope \citep{Vegetti et al. b, Vegetti et al. c}. 

By mapping extended macrolensed sources with milliarcsecond or sub-milliarcsecond resolution using Very Long Baseline Interferometry (VLBI) techniques at radio wavelengths, substructures at even lower masses can in principle be detected. Such objects may introduce kinks and bends in multiply-imaged quasar jets \citep{Wambsganss & Paczynski,Metcalf & Madau} and one detection of a $\sim 10^5$--$10^7 M_\odot$ object has already been claimed using this technique \citep{Metcalf}. In this situation, the lensing effects produced by halo substructures can be separated from intrinsic morphological features in jets, since the latter would be reproduced in all macroimages whereas dark matter clumps in the halo of the lens would affect each macroimage differently. Similar methods for exploiting the lensing effects produced by halo substructures on scales of $\sim 100$ milliarcseconds down to $\sim 0.01$ milliarcseconds have also been explored by \citet{Yonehara et al.,Inoue & Chiba a,Inoue & Chiba b,Inoue & Chiba c, Hisano et al.,Ohashi et al.,Riehm et al.} and \citet{Hezaveh et al.}. However, effects of this type tend to be sensitive to the density profiles of substructures, and may be undetectable for all but the very densest, most extreme forms of substructure \citep{Zackrisson et al.}.

Here, we use lensing simulations to explore the prospects of using macrolensed quasar jets observed at sub-milliarcsecond resolution, in searches for standard CDM subhalos, ultracompact minihalos and primordial black holes within the main lens. These different forms of substructure are described, along with previous constraints on such objects, in Sect.~\ref{substructures}. The details of our simulations and assumptions are covered in Sect.~\ref{simulations}. In Sect.~\ref{results}, we present our results and in Sect.~\ref{discussion} we discuss some lingering issues with the adopted technique. Sect.~\ref{summary} summarizes our findings.

\section{Different forms of halo substructure}
\label{substructures}
\subsection{Standard CDM subhalos}
\label{standard CDM}
At $z=0$, the CDM scenario predicts the existence of dark matter halos with masses ranging from $\sim 10^{15}\ M_\odot$ down to the cutoff in the density fluctuation spectrum, which is set by the detailed properties of the CDM particles. For many types of WIMPs, this cut-off lies somewhere in the range $\sim 10^{-11}$--$10^{-3} M_\odot$ \citep{Bringmann}. Alternative models involving superweakly-interacting particles (super-WIMPS), MeV mass dark matter or a long-range interaction between dark matter particles may place the cutoff as high as $\sim 10^3$--$10^{10} \ M_\odot$ \citep{Hisano et al.,Hooper et al.,van den Aarssen et al.}, although the upper end of this range may be in conflict with the apparent masses of the lightest dwarf galaxies \citep[$\sim 10^6\ M_\odot$;][]{Geha et al.}.

As these low-mass halos merge to form more massive ones, some temporarily survive in the form of subhalos within the larger halos. N-body simulations indicate that the subhalos within a galaxy-sized CDM halo should follow a mass function of the type:
\begin{equation}
\frac{\mathrm{d}N}{\mathrm{d}M_\mathrm{sub}}\propto M_\mathrm{sub}^{-\alpha},
\label{subhalo mass function}
\end{equation}
with $\alpha\approx 1.9$ \citep{Springel et al.,Gao et al.}. The relative contribution from subhalos with mass $M\gtrsim 10^5\ M_\odot$ to the dark matter surface mass density at the typical location of macroimages in a galaxy-mass lens is $f_\mathrm{sub}\approx 0.002$ \citep{Xu et al. b}, albeit with a large scatter \citep{Chen et al.}. 

The density profiles of isolated field halos in CDM simulations can be reasonably well described by Navarro, Frenk \& White (\citealt{NFW}; hereafter NFW) profiles:
\begin{equation}
\rho(R)=\frac{\rho_\mathrm{i}}{(R/R_\mathrm{S})(1+R/R_\mathrm{S})^{2}}, 
\label{NFW}
\end{equation}
where $R_\mathrm{S}$ is the characteristic scale radius of the halo. The slope of the inner density cusp ($\beta = \mathrm{d}\ln \rho /\mathrm{d}\ln r $) in this profile is $\beta = -1$, and this makes it difficult for NFW halos in the dwarf-galaxy mass range to produce millilensing effects of the type we are considering in this paper. Typically, cusp slopes obeying $\beta\lesssim -1.5$ would be required \citep{Zackrisson et al.}. Later work has shown that models with inner cusp slopes that become progressively shallower towards the centre provide even better fits to isolated halos in CDM simulations, eventually reaching inner slopes of $\beta \geq -1$ \citep[e.g.][]{Navarro et al. 10}. In the context of detecting millilensing effects from low-mass halos, this just makes matters worse, since the central density is reduced.

Once a halo falls into the potential well of a larger halo and becomes a subhalo, it is stripped of material -- primarily from its outskirts -- due to interactions with its host halo and with other subhalos. This alters the density profile of the subhalo compared to an isolated halo of the same mass \citep[e.g.][]{Hayashi et al.,Kazantzidis et al.}, but these modifications tend to diminish rather than enhance the ability of a CDM subhalo to produce detectable millilensing effects \citep{Zackrisson et al.}. 

To demonstrate that standard CDM subhalos do not provide a significant ``background'' of millilensing events in the observational situation that we consider, we have therfore adopted NFW profiles for these objects, since this results in an overoptimistic estimate on the millilensing effects that standard CDM subhalos are likely to produce. Even then, the chances of detecting millilensing effects from these objects turn out to be negligible in the observational situations we are considering. 

To derive the $R_\mathrm{S}$ values of our NFW subhalo profiles, we adopt the mass-dependent concentration parameters $c=R_\mathrm{vir}/R_\mathrm{S}$ from either \citet{Bullock et al.} or \citet{Maccio et al. a}, where $R_\mathrm{vir}$ is the virial radius of the halo. Since both of these recipes predict higher concentration parameters for low-mass halos, and since more centrally concentrated profiles (i.e. profiles with higher $c$) are more efficient in producing millilensing effects, we calculate the subhalo concentration parameters based on their current masses rather than the masses they had prior to becoming subhalos. Since nearly all subhalos have lost considerable amounts of material \citep[e.g.][]{Vale & Ostriker}, this also results in overly optimistic millilensing properties. 

\subsection{Intermediate-mass black holes}
Intermediate-mass black holes (IMBHs; here assumed to have masses $\sim 10^{3}$--$10^6\ M_\odot$) may either form primordially (typically when the Universe is $\ll 1$ s old), or due to the collapse of baryonic objects later on. The primordial variety could in principle comprise a substantial fraction of the dark matter, although a host of observational constraints makes this seem unlikely \citep[for a recent compilation, see][]{Carr et al.}.

The strongest upper limits on the cosmological density of primordial black holes in the $10^{3}$--$10^5\ M_\odot$ mass range come from the effect that accretion onto these objects would have on the cosmic microwave background radiation \citep{Ricotti et al.}. Primordial black holes with masses $\sim 10^3$--$10^4\ M_\odot$ are also strongly constrained by the effects of gravity waves on pulsar timing measurements \citep{Saito & Yokoyama,Carr et al.}, and at $M\gtrsim 10^4\ M_\odot$ by dynamical constraints \citep{Carr & Sakellariadou}. Using a technique first proposed by \citet{Kassiola et al.}, \citet{Wilkinson et al.} moreover used the absence of millilensing effects in non-macrolensed radio sources to place upper limits on IMBHs at $M\gtrsim 10^5\ M_\odot$. It has, however, been argued that some of these constraints may be sidestepped under certain circumstances, and that both the size evolution of early-type galaxies \citep{Totani et al.} and entropy considerations \citep{Frampton et al.} favour scenarios in which essentially all of the dark matter is in the form of $\sim 10^5\ M_\odot$ primordial black holes. 
 
Intermediate-mass black holes that were not produced primordially may instead either form as the end products of very massive population III stars \citep[e.g.][]{Madau & Rees}, through the direct collapse of gas in small halos at high redshift \citep[e.g.][]{Begelman et al.} or the collapse of dense star clusters \citep[e.g.][]{Devecchi & Volonteri,Davies et al.}. Such IMBHs may now be hiding in globular clusters \citep[e.g.][]{Vesperini et al.}, in satellite galaxies \citep{van Wassenhove et al.}, or be freely floating in the halos of galaxies \citep[e.g.][]{Micic et al.}. There is indeed some evidence for IMBHs in globular clusters \citep[e.g.][]{Noyola et al.}, and IMBHs may also explain some of the ultraluminous X-ray sources detected in other galaxies \citep[e.g.][]{Feng & Soria,Webb et al.}. However, since only a small fraction of the cosmic baryons can be locked up in these non-primordial IMBHs, their relative contributions to the halo masses of galaxies are typically expected to be small \citep[$f_\mathrm{IMBH}\lesssim 10^{-5}$; e.g. ][]{Islam et al.,Kawaguchi et al.}.

When simulating the potential millilensing effects of IMBHs, we treat the surface mass density fraction $f_\mathrm{IMBH}$ in IMBHs at the position of the macroimages as a free parameter, and for simplicity assume all IMBHs to have the same mass. In the case where the number density profile of IMBHs has the same shape as the density profile of the dark halo, $f_\mathrm{IMBH}$ also corresponds to the halo mass fraction in IMBHs. This is expected to be the case for primordial black holes, even if they constitute no more than a small fraction of the dark matter, since such objects behave just like CDM particles in N-body simulations. In the case of IMBHs formed through baryonic processes, the number density profile of IMBHs may well deviate significantly from the overall dark matter profile, and an $f_\mathrm{IMBH}$ estimate obtained from strong lensing observations cannot directly be interpreted as halo mass fraction without further constraints on the baryon distribution within the lens. 

In principle, primordial black holes may over time accrete substantial amounts of dark matter and develop dark matter halos of their own \citep{Mack et al.}, similar to the ultracompact minihalos discussed in Sect.~\ref{UCMH_section}. IMBHs forming through the collapse of pop III stars in minihalos may also be surrounded by their own, highly contracted dark matter halos \citep[][]{Sandick et al.}. Such compound objects are expected to have lensing properties intermediate between IMBHs and ultracompact minihalos, but are not treated in detail in our simulations. 

\subsection{Ultracompact minihalos}
\label{UCMH_section}
Primordial density perturbations with $\Delta\rho/\rho\equiv\delta\lesssim0.3$ are too small to form primordial black holes as they enter the horizon.  Those with $\delta\gtrsim10^{-3}$ may nonetheless still be large enough to seed the formation of ultracompact minihalos \citep[UCMHs;][]{Berezinsky et al. a, Berezinsky et al. b, Berezinsky et al. c, Berezinsky et al. d,Ricotti & Gould,Berezinsky et al. e,Bringmann et al. a}.  Such perturbations might be produced during phase transitions, around topological defects, or in the primordial spectrum of perturbations from inflation.  The dark matter contained in these perturbations would collapse into UCMHs shortly after matter-radiation equality, via radial infall from a universally cold, smooth cosmological background.  This radial collapse would leave UCMHs with much steeper central density profiles than standard CDM halos \citep{Ricotti & Gould}.

If dark matter exists in the form of self-annihilating WIMPs, UCMHs would be gamma-ray emitters, and strong limits on their cosmological density have already been derived from the effect that this would have on Fermi-LAT source identifications, the diffuse gamma-ray background and cosmic reionization \citep{Scott & Sivertsson,Josan & Green,Lacki & Beacom,Berezinsky et al. e,Berezinsky et al. f,Yang et al. a,Yang et al. b,Zhang,Bringmann et al. a,Shandera et al.}; similarly if dark matter decays rather than annihilating \citep{Yang et al. c}. If dark matter does not annihilate or decay, UCMHs in the $\sim$$10^{-2}$--$10^2\,M_\odot$ range may still be detectable in the future by their astrometric lensing effects on Milky Way stars \citep{Li et al.}.

Here, we explore to what extent submilliarcsecond observations of macrolensed jets would be able to constrain the properties of UCMHs. Because dark matter self-annihilation would reduce the central density of UCMHs \citep[e.g.][]{Scott & Sivertsson}, UCMHs made out of self-annihilating WIMPs would not be efficient millilenses. We therefore focus on UCMHs made out of non-annihilating dark matter (e.g. asymmetric dark matter, axions, sterile neutrinos).

Radial infall leads to a density profile $\rho\propto r^{-2.25}$, slightly steeper than the $\rho\propto r^{-2}$ profile of a singular isothermal sphere (often used to model lensing by the inner regions of large galaxies).  The dark matter profile in a UCMH \citep[see][ for a detailed discussion]{Ricotti & Gould,Bringmann et al. a} is given by
\begin{equation}
\label{UCMHprofile}
\rho(r,z)=\frac{3f_\mathrm{CDM} M_\mathrm{UCMH}(z)}{16\pi R_\mathrm{UCMH}(z)^\frac34r^\frac94},
\end{equation}
where $f_\mathrm{CDM}$ is the cosmological fraction of matter in CDM, $M_\mathrm{UCMH}(z)$ indicates the UCMH mass at redshift $z$, and
\begin{equation}
\label{UCMHradius}
\frac{R_\mathrm{UCMH}(z)}{\mathrm{pc}}=0.019\left(\frac{1000}{z+1}\right)\left(\frac{M_\mathrm{UCMH}(z)}{M_\odot}\right)^\frac13,
\end{equation}
is the UCMH radius, defined as the distance within which the density is at least twice that of the cosmological background.  Following matter-radiation equality at $z_{\rm eq}$, a UCMH born from an initial dark matter overdensity of mass $M_i$ accretes both dark and baryonic matter as
\begin{equation}
\label{Mz}
M_{\rm UCMH}(z)= \frac{z_{\rm eq}+1}{z+1}M_i\,.
\end{equation}
This accretion presumably cuts out when the cosmological background is no longer smooth, i.e. when the first substantial structure formation occurs and the smallest star-forming minihalos appear.  In this case, present-day UCMH masses and radii can be obtained by setting $z\sim10$ in Eqs. (\ref{UCMHprofile}) and (\ref{UCMHradius}), so that 
\begin{align}
M^0_\mathrm{UCMH} &\equiv M_\mathrm{UCMH}(z\lesssim10) = M_\mathrm{UCMH}(z=10),\\R^0_\mathrm{UCMH} &\equiv R_\mathrm{UCMH}(z\lesssim10) = R_\mathrm{UCMH}(z=10).
\end{align}

However, the finite temperature of the smooth cosmological background from which UCMHs accrete softens the density profile in the innermost region, due to conservation of angular momentum.  This can be conservatively modelled as a cutoff at some inner radius $r_{\rm min}$, inside which one assumes the density to be constant.  Following previous work \citep{Bringmann et al. a}, we adopt this strategy here, taking a flat density profile within
\begin{equation}
\label{rmin}
\frac{r_\mathrm{min}}{R^0_\mathrm{UCMH}} \approx 2.9\times10^{-7}\left(\frac{1000}{z_c+1}\right)^{2.43}\left(\frac{M^0_\mathrm{UCMH}}{M_\odot}\right)^{-0.06}\,.
\end{equation}
Here $z_c$ refers to the redshift of UCMH collapse (the point at which the growth of the matter overdensity becomes non-linear); we adopt $z_c=1000$, also in line with earlier work \citep{Ricotti & Gould, Bringmann et al. a}.

\section{Lensing simulations}
\label{simulations}
To simulate the effects of dark halo substructure on the morphologies of macrolensed jets, we use a numerical scheme similar to that developed by \citet{Metcalf & Madau}. An extended source is assumed to be multiply-imaged by a foreground galaxy, and the lens equation is used to determine the lens plane positions of the corresponding macroimages. A small region around each such macroimage is then populated with randomly distributed dark halo substructures and simulated in greater detail. The deflection angles (with contributions both from substructures and the macrolens) are computed for every pixel within this region and converted into a numerical surface brightness map of the macroimage. These maps are initially generated with a very fine pixel scale, but are then convolved with a Gaussian filter to match the finite resolution  of the VLBI arrays we consider. Both the resolution and the intrinsic source dimensions are determined by the frequency at which we assume the jets to be observed, as described in Sect.~\ref{VLBI} and~\ref{source size}. 
 
The macrolens is modelled as a singular isothermal sphere \citep[as appropriate for early-type galaxies acting as strong lenses; e.g.][]{Rusin et al. b} at $z_\mathrm{l}=0.5$, with line-of-sight velocity dispersion $\sigma_\mathrm{v}=240$ km s$^{-1}$, giving a mass of $\sim 10^{13} \ M_\odot$ within the virial radius, and two macroimages with separation $\approx 2\arcsec$. We furthermore adjust the alignment of the source and main lens to ensure a macrolens magnification that is not unrealistically high. The simulations presented in this paper are all based on a lens-source configuration that in the absence of substructure would give magnifications $\mu_1\approx 10$ and $\mu_2\approx 8$ for the two macroimages. All simulations are based on a $\Lambda$CDM cosmology with $H_0=70$ km s$^{-1}$ Mpc$^{-1}$, $\Omega_\mathrm{M}=0.27$, $\Omega_\Lambda=0.73$. 

When distributing halo substructures within the simulated region, we for simplicity assume that the surface mass density across the macroimage is completely dominated by dark matter. While this assumption may be violated in multiply-imaged systems where one of the macroimages happens to lie very close to the lensing galaxy, this nonetheless seems to be a fair approximation in the majority of cases \citep[e.g.][]{Bate et al.,Pooley et al.}. In the case of intermediate mass black holes and ultracompact minihalos, we moreover assume that their number densities trace that of the dark matter. The surface number density of such subtructures then simply depends on their relative contribution to dark matter $\Omega_\mathrm{sub}/\Omega_\mathrm{CDM}$ and their mass distribution. These dark matter fractions in intermediate-mass black holes and ultracompact minihalos are referred to as $f_\mathrm{IMBH}$ and $f_\mathrm{UCMH}$ respectively. Since detailed predictions for the mass distribution of IMBHs and UCMHs are highly model-dependent, we assume all such objects to have the same mass, which we then vary to explore what parts of the ($f_\mathrm{IMBH}$,$M_\mathrm{IMBH}$) or ($f_\mathrm{UCMH}$,$M_\mathrm{UCMH}$) parameter space that a given set of observations would be able to probe.

In the case of standard CDM subhalos, we adopt the mass distributions inferred from either simulations or observations. As discussed in Sect.~\ref{standard CDM}, current simulations suggest $f_\mathrm{NFW}=0.002$ at the typical positions of macroimages in galaxy-sized dark halos, and the subhalo mass function given by Eq. (\ref{subhalo mass function}). However, since the recent lensing detections of subhalos by \citet{Vegetti et al. b,Vegetti et al. c} hint at a flatter mass function and a mass fraction that is an order of magnitude higher, we also explore the consequences of setting $f_\mathrm{NFW}=0.03$ and changing the mass function slope of Eq.~(\ref{subhalo mass function}) to $\alpha=1.1$.

\begin{figure*}
\includegraphics[scale=0.29]{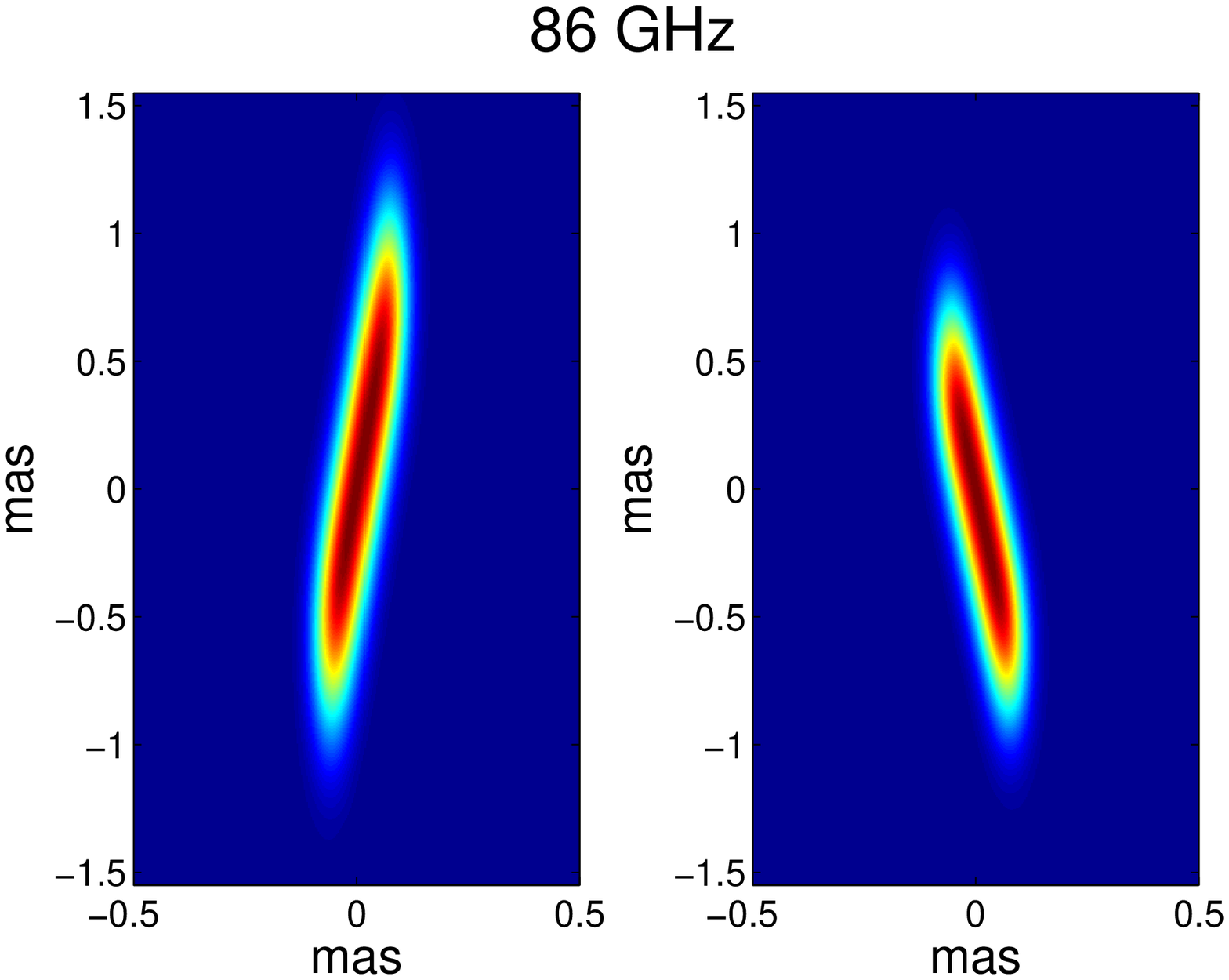}
\includegraphics[scale=0.29]{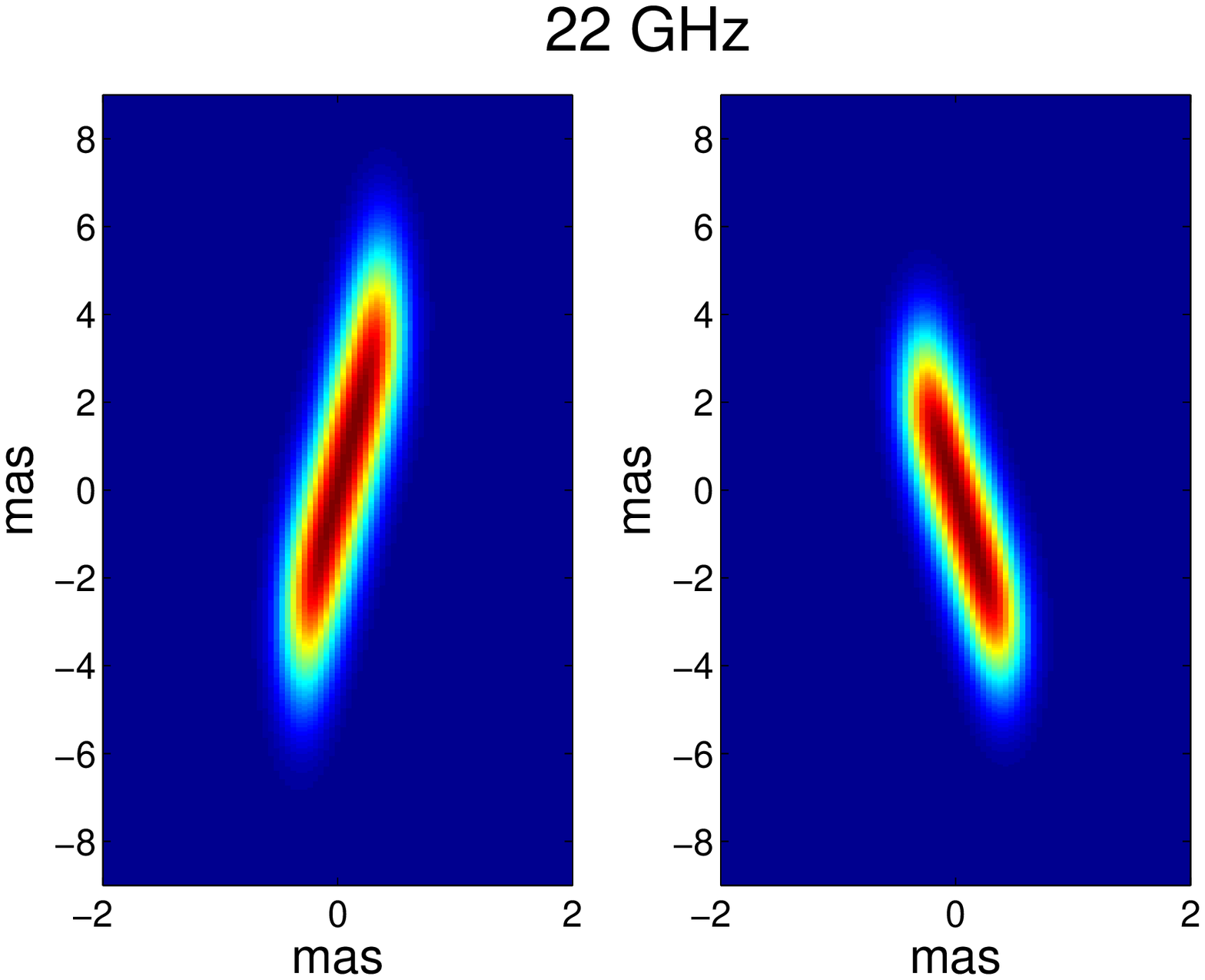}
\includegraphics[scale=0.28]{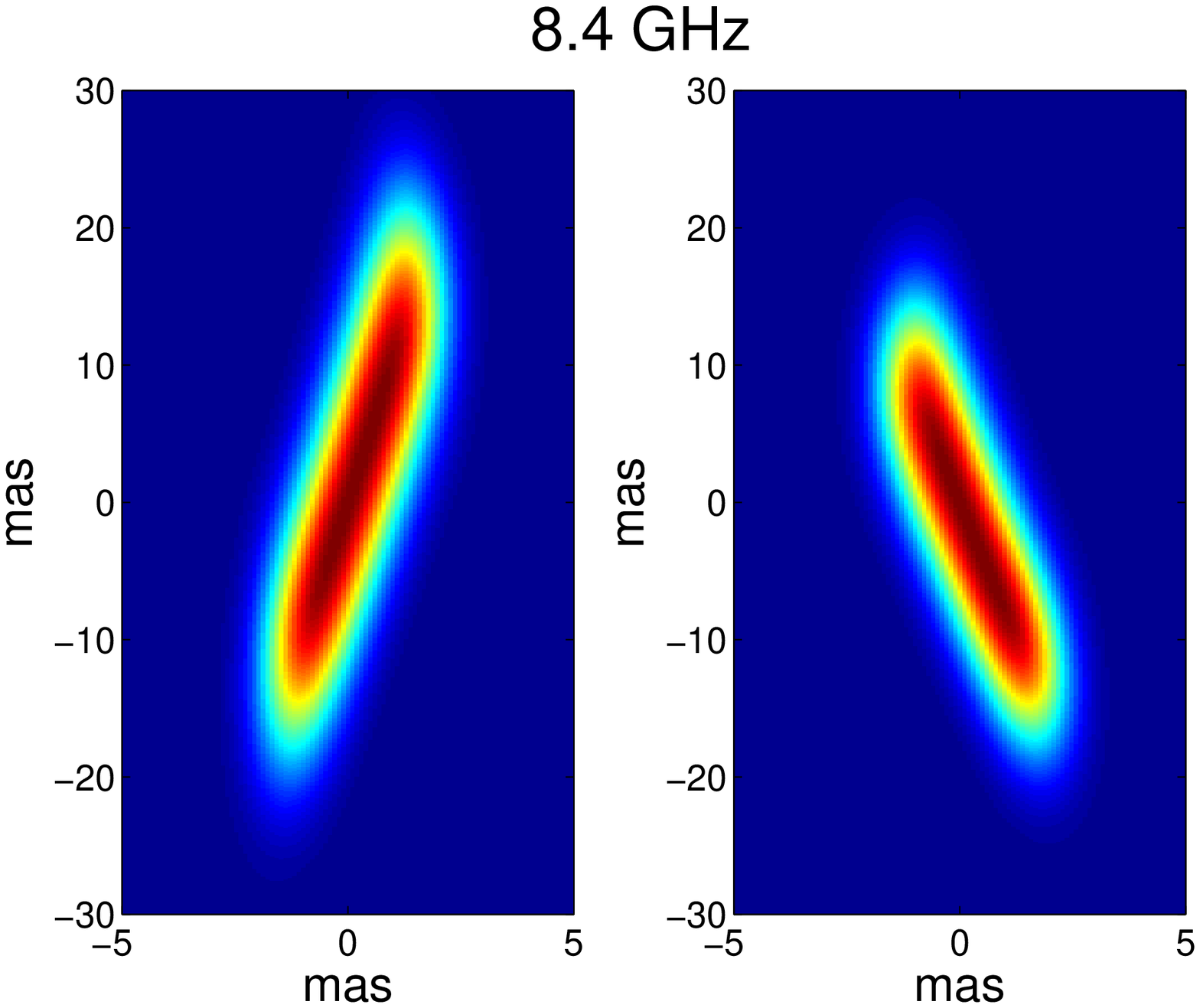}
\\
\includegraphics[scale=0.29]{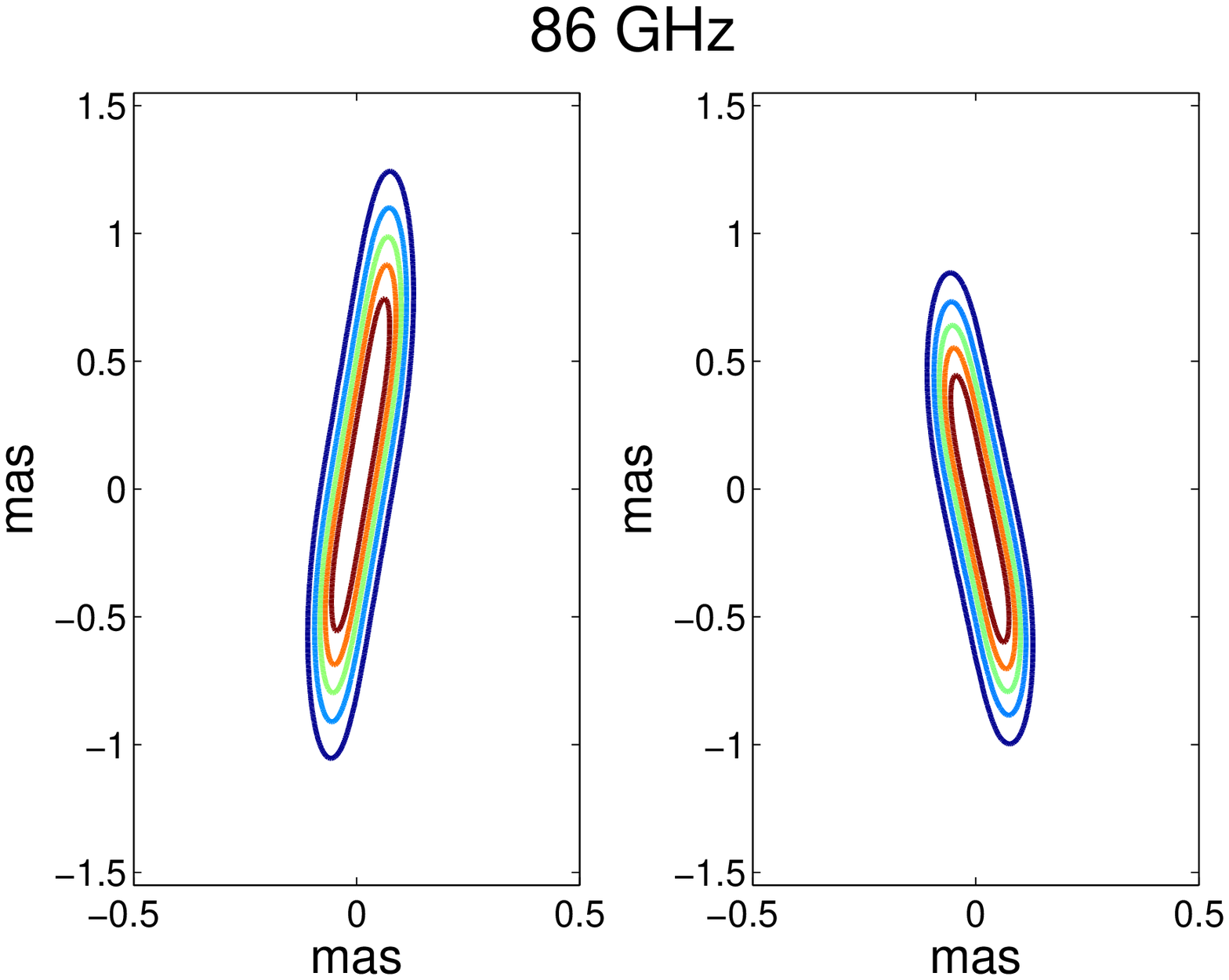}
\includegraphics[scale=0.29]{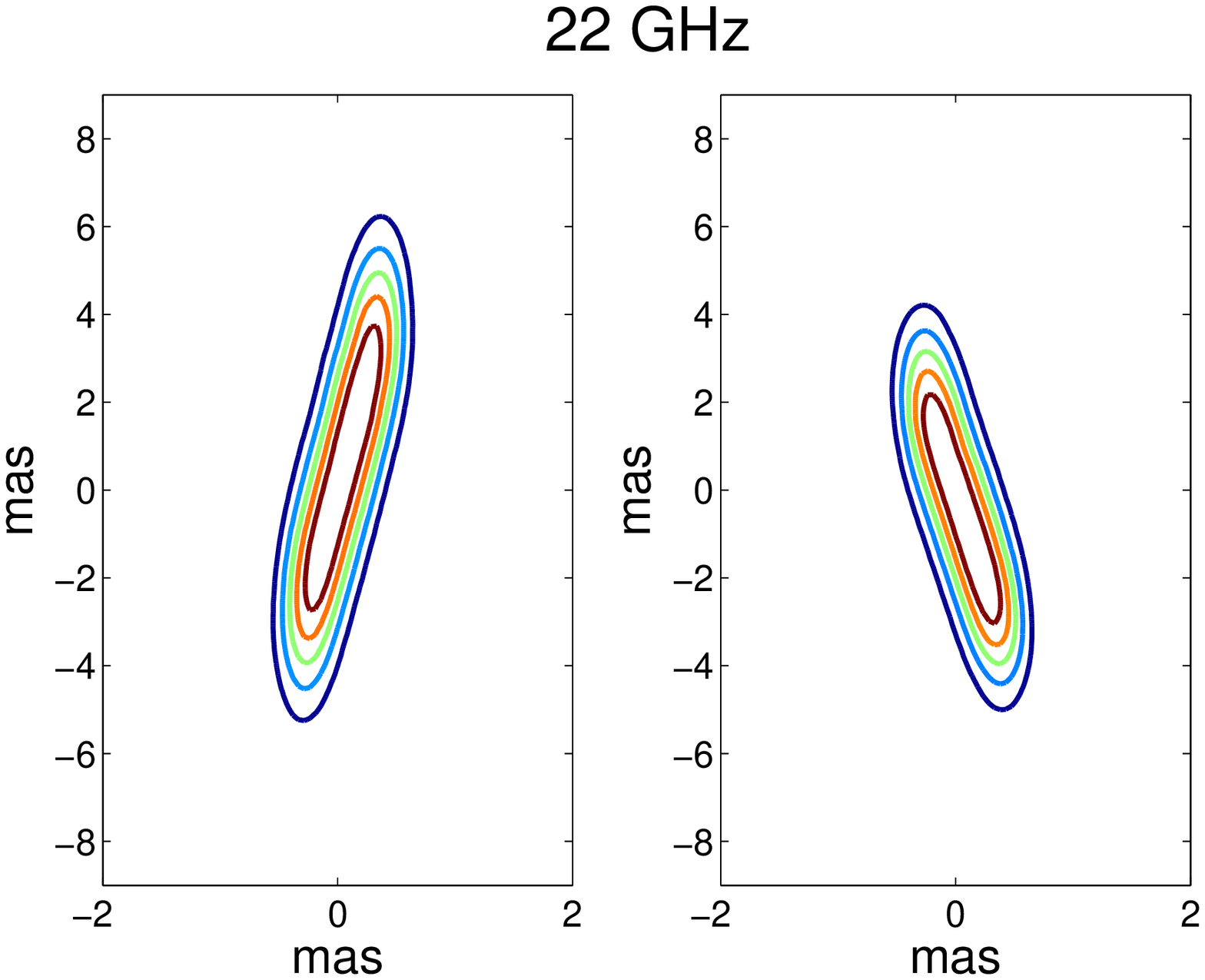}
\includegraphics[scale=0.28]{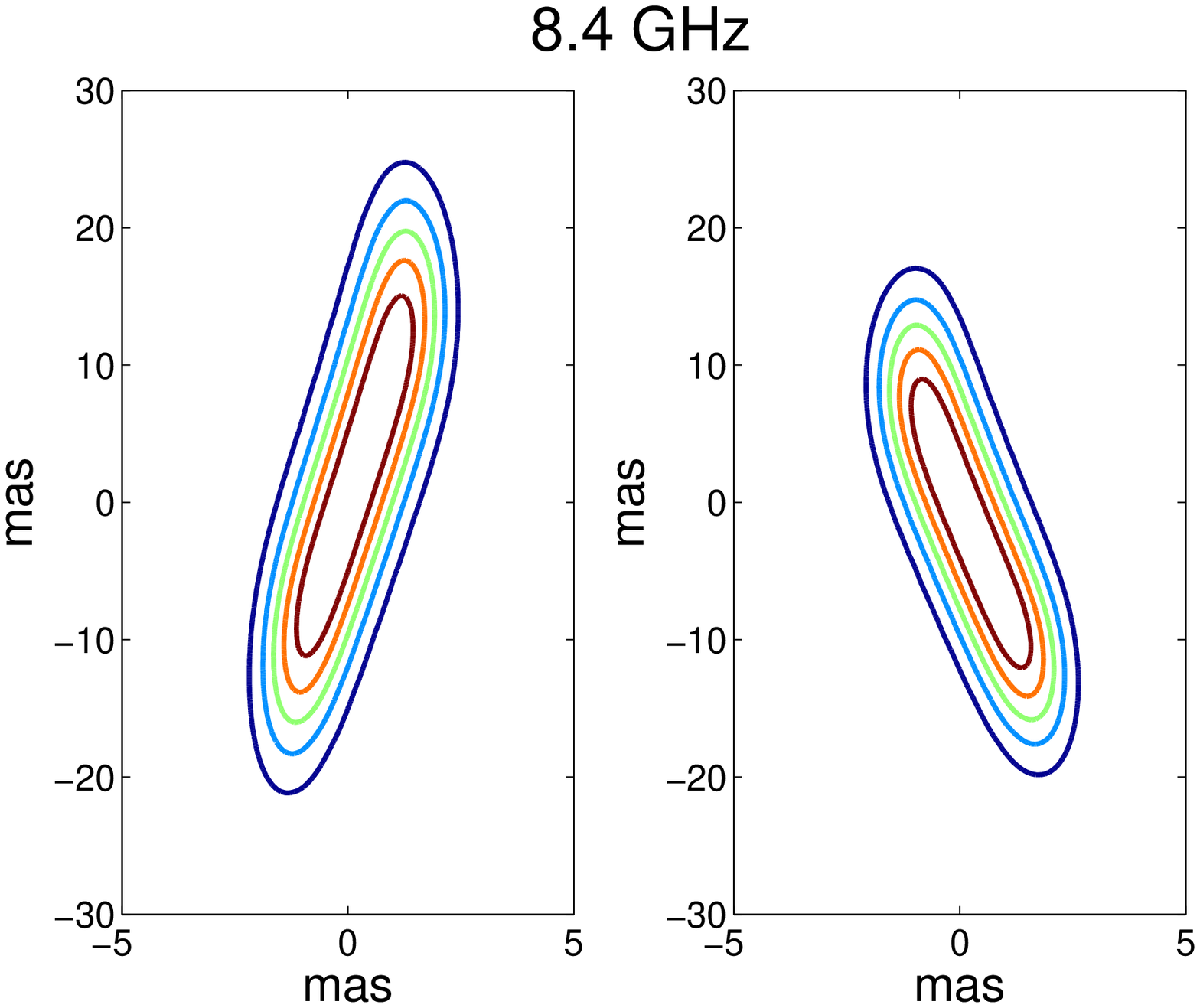}
\caption{Simulated radio maps of strongly lensed quasar jets at 86 GHz, 22 GHz and 8.4 GHz, respectively from left to right (source sizes $2\times 0.5$ pc, $10\times 2.5$ pc and $40\times 10$ pc), subject to macrolensing by the main lens. The two subplots of each image show the two macroimages of the source. The bottom row contains the contour representations of the macroimages in the upper row, with the outermost contours corresponding to $\approx 10\%$ of the peak flux in these images. Please note the different scales of the images at the three frequencies.}
\label{imagemap_nosub_all}
\end{figure*}

\subsection{VLBI observations}
\label{VLBI}
A number of macrolensed radio jets are already known and have been observed using VLBI techniques \citep[e.g.][]{Garrett et al.,King et al.,Ros et al.,Rusin et al. a,Biggs et al.,York et al.}, although typically not with arrays capable of resolving submilliarcsecond scales. Designing a survey aimed to search for small-scale distortions in targets like these does, however, also involve other considerations than just the resolution. The frequency at which one chooses to observe these jets limits the resolution at which the jets can be mapped using suitable VLBI arrays, but also affects the intrinsic source size \citep[e.g.][]{Torniainen et al.}. To identify the observational strategy that maximizes the scientific output in terms of detection prospects for dark halo substructure, we here consider observations at three different frequencies, each using a different VLBI array:
\begin{itemize}
\item Observations at 8.4 GHz using the global array, including the European VLBI Network (EVN\footnote{http://www.evlbi.org/}) and the Very Long Baseline Array (VLBA\footnote{http://www.vlba.nrao.edu/}), giving a resolution of $\approx 0.7$ milliarcseconds
\item Observations at 22 GHz using the EVN, giving a resolution of $\approx 0.3$ milliarcseconds 
\item Observations at 86 GHz using the full Atacama Large Millimeter (ALMA\footnote{http://www.almaobservatory.org/}) array (66 antennas) connected to the global 3 mm array\footnote{http://www.mpifr-bonn.mpg.de/div/vlbi/globalmm/}, giving a resolution of $\approx 0.05$ milliarcsec. This observing mode is not available at the current time, but is likely to come on line in a few years. 
\end{itemize}  
These arrays also have different sensitivities, which constrains the numbers of potential targets and also the apparent lengths of the jets. However, since we are simulating the effects of generic sources rather than individual targets, this is not addressed in our current simulations. 

\subsection{Source size and morphology}
\label{source size}
We assume the source to be an intrinsically straight jet with a
2-dimensional Gaussian surface brightness profile and length 40, 10 and 2 pc at 8.4,
22 and 86 GHz, respectively, and a width that is a quarter of the
length. These sizes are in a reasonable agreement with source size
estimates presented in \citet{Torniainen et al.} and the jet lengths
calculated from the MOJAVE sample \citep{Lister et
  al.}\footnote{http://www.physics.purdue.edu/MOJAVE/}.
  
The source morphology and length-to-width ratio are mainly adopted for illustrative
purposes. The limits presented in Sect.~\ref{results} are fairly insensitive
to these assumptions, and depend mainly on the intrinsic jet area, as further discussed in Sects.~\ref{size} and ~\ref{sbp}. At fixed
angular resolution, a larger jet results in a stronger constraint
whereas a smaller jet makes the constraints weaker. The results of
Sect.~\ref{results} can therefore be rescaled to accommodate other jet
sizes. In broad terms, our assumptions on the source sizes are similar
to those used by \citet{Inoue & Chiba a}. 
   
\section{Results}
\label{results}
\begin{figure*}
\includegraphics[scale=0.4]{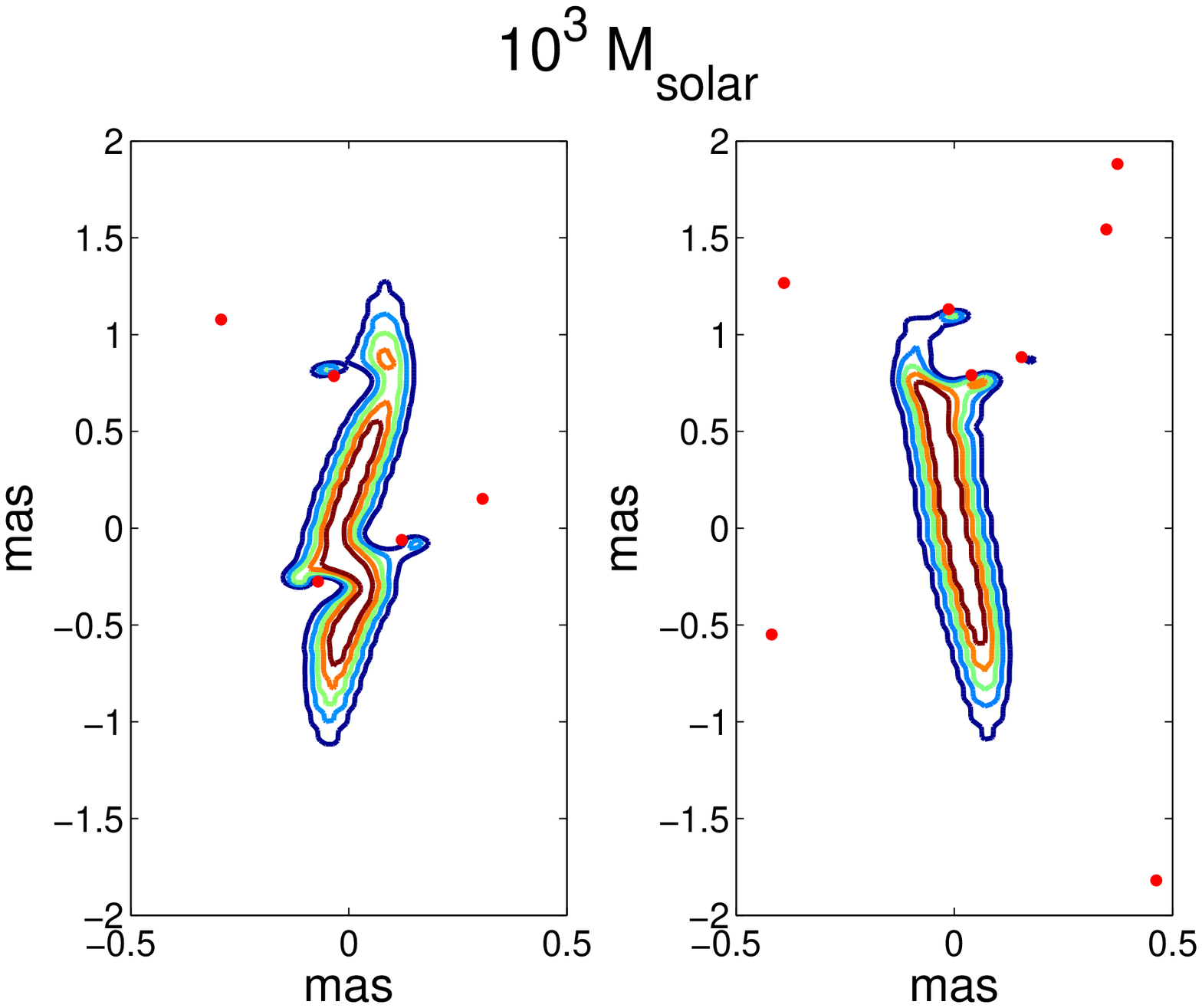}
\includegraphics[scale=0.4]{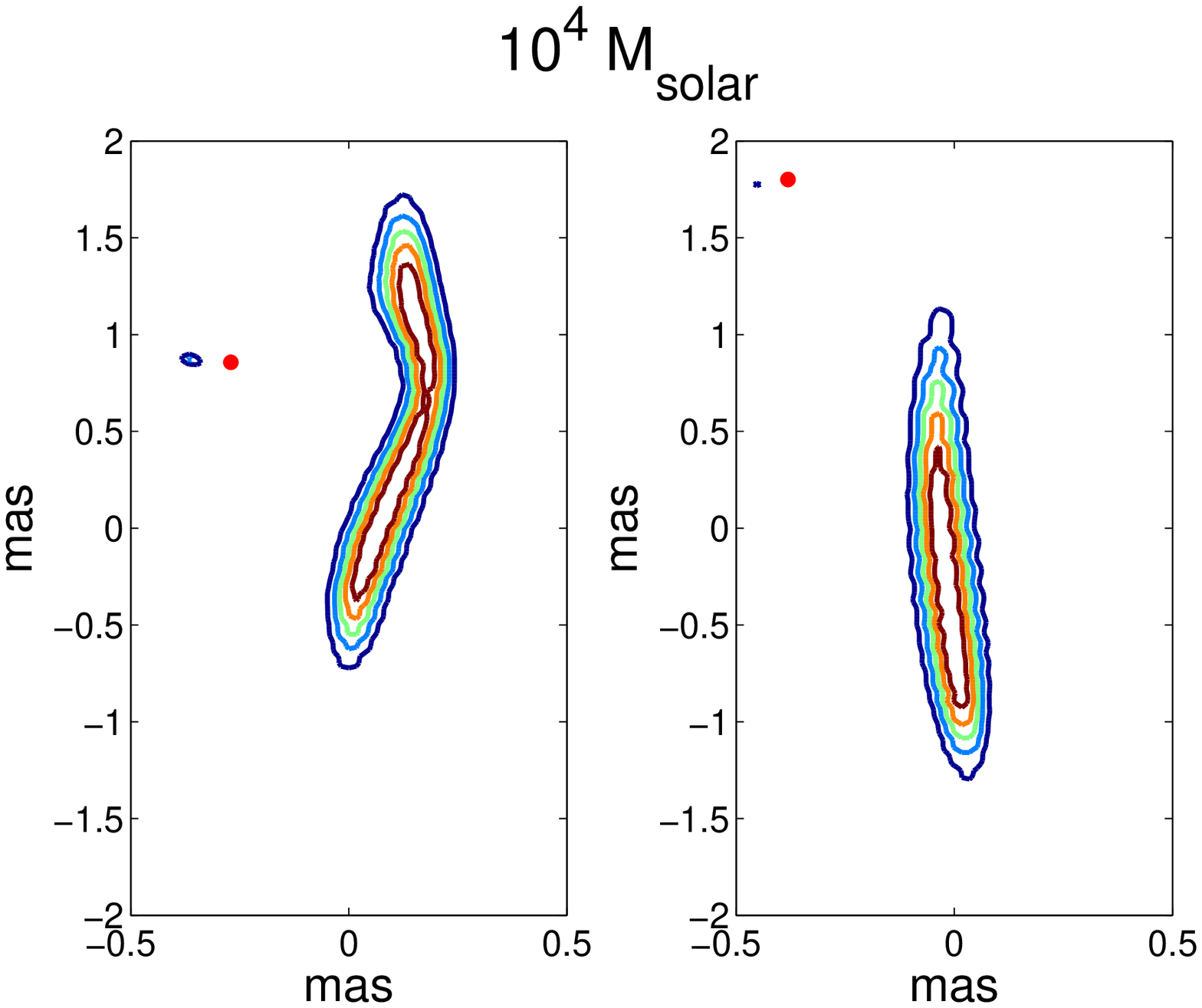}\\
\includegraphics[scale=0.4]{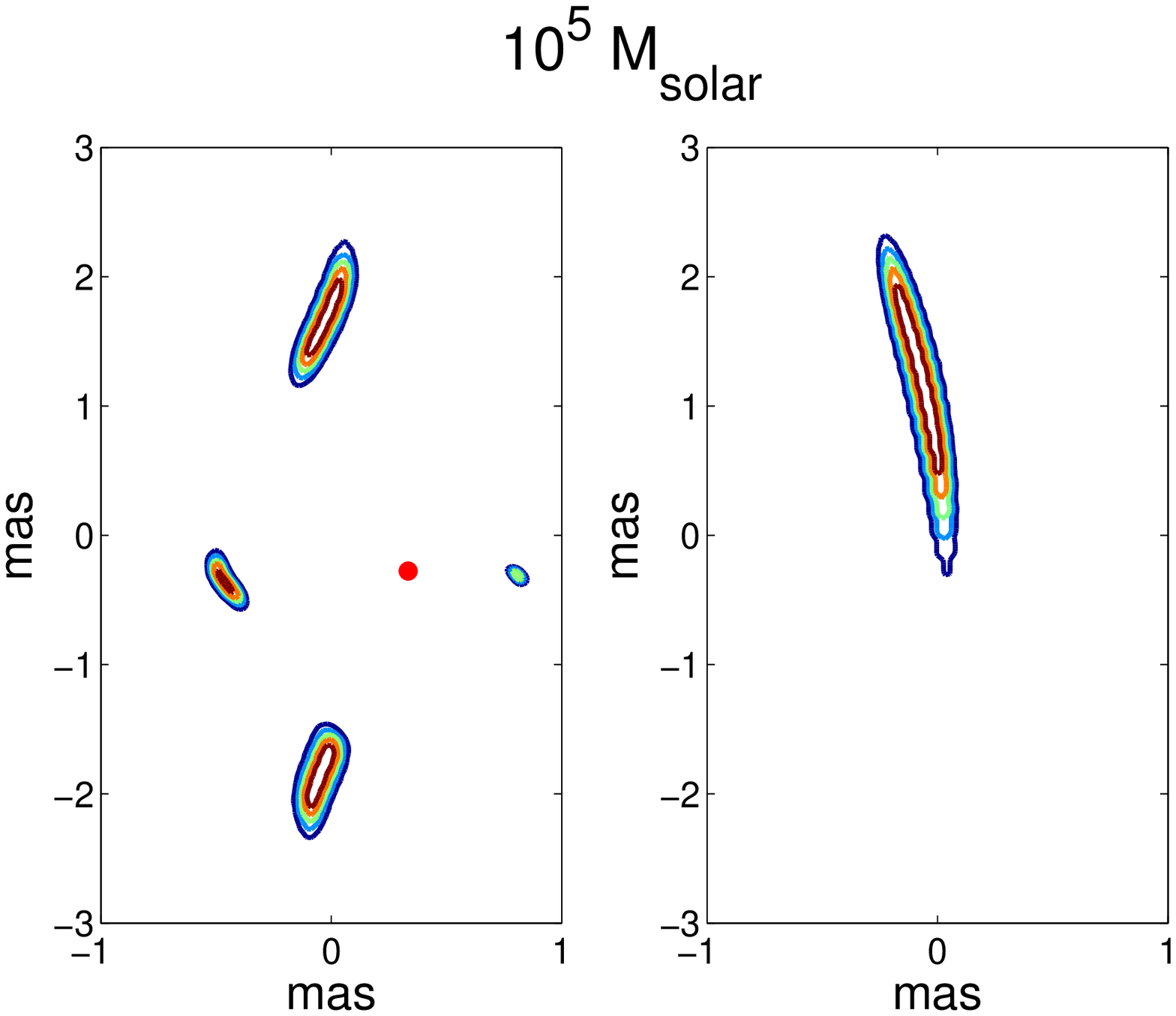}
\includegraphics[scale=0.4]{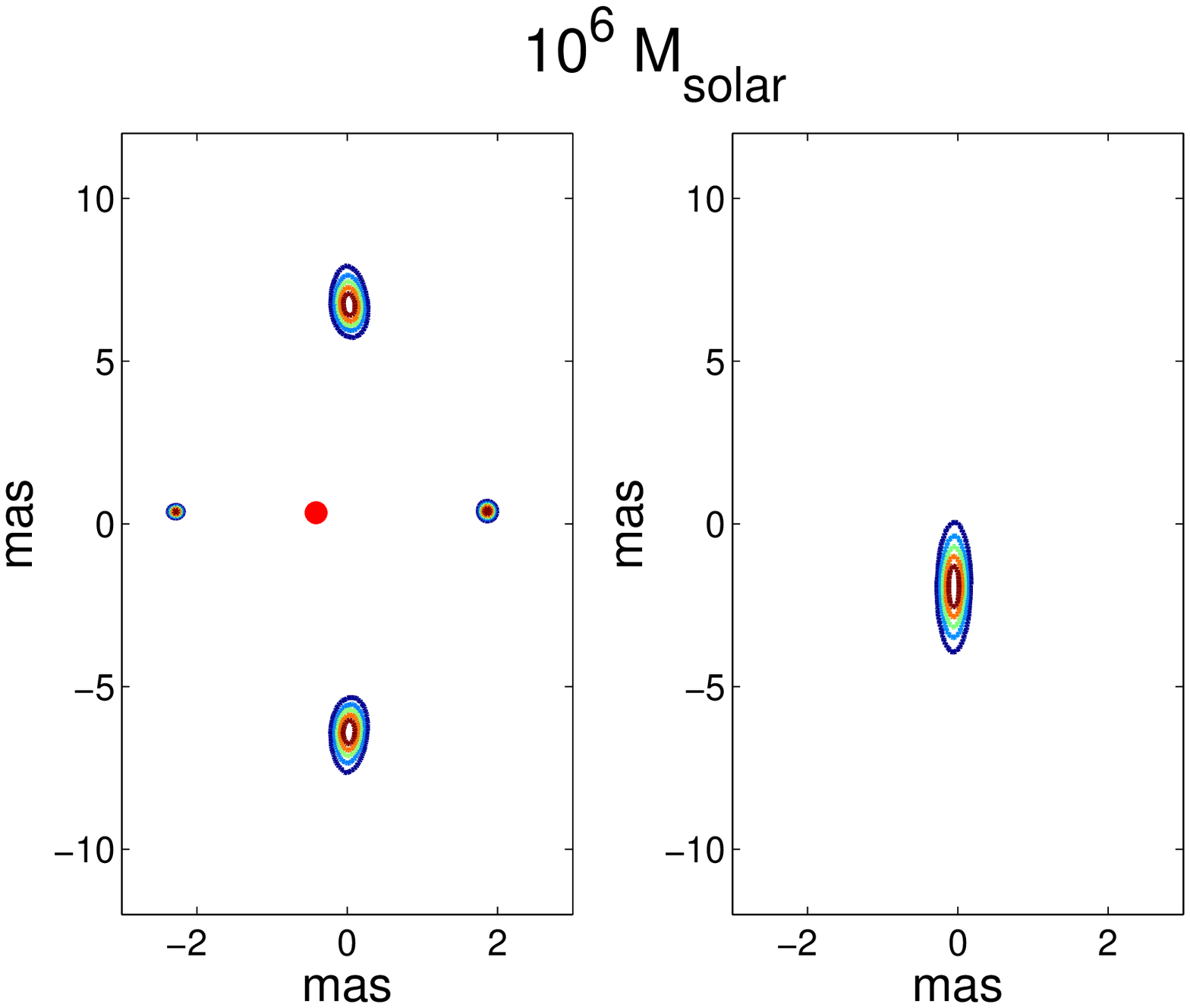}
\caption{Examples of simulated radio maps of a macrolensed quasar jet at 86 GHz (source size $2\times 0.5$ pc and resolution 0.05 milliarcseconds), subject to millilensing distortions by IMBHs with $f_\mathrm{IMBH}=0.02$ in the halo of the main lens. Each image pair represents the two macroimages from Fig.~\ref{imagemap_nosub_all}, distorted by millilensing effects from IMBHs with either $M_\mathrm{IMBH}=10^3$, $10^4$, $10^5\ M_\odot$ or $10^6\ M_\odot$. The positions of the IMBHs are indicated by red dots. The slight macroimage distortions and displacements seen in right panels for the $10^5$ or $10^6\ M_\odot$ cases are produced by IMBHs just outside the plotted region.}
\label{imagemap_IMBH_86GHz}
\end{figure*}

\begin{figure*}
\includegraphics[scale=0.3]{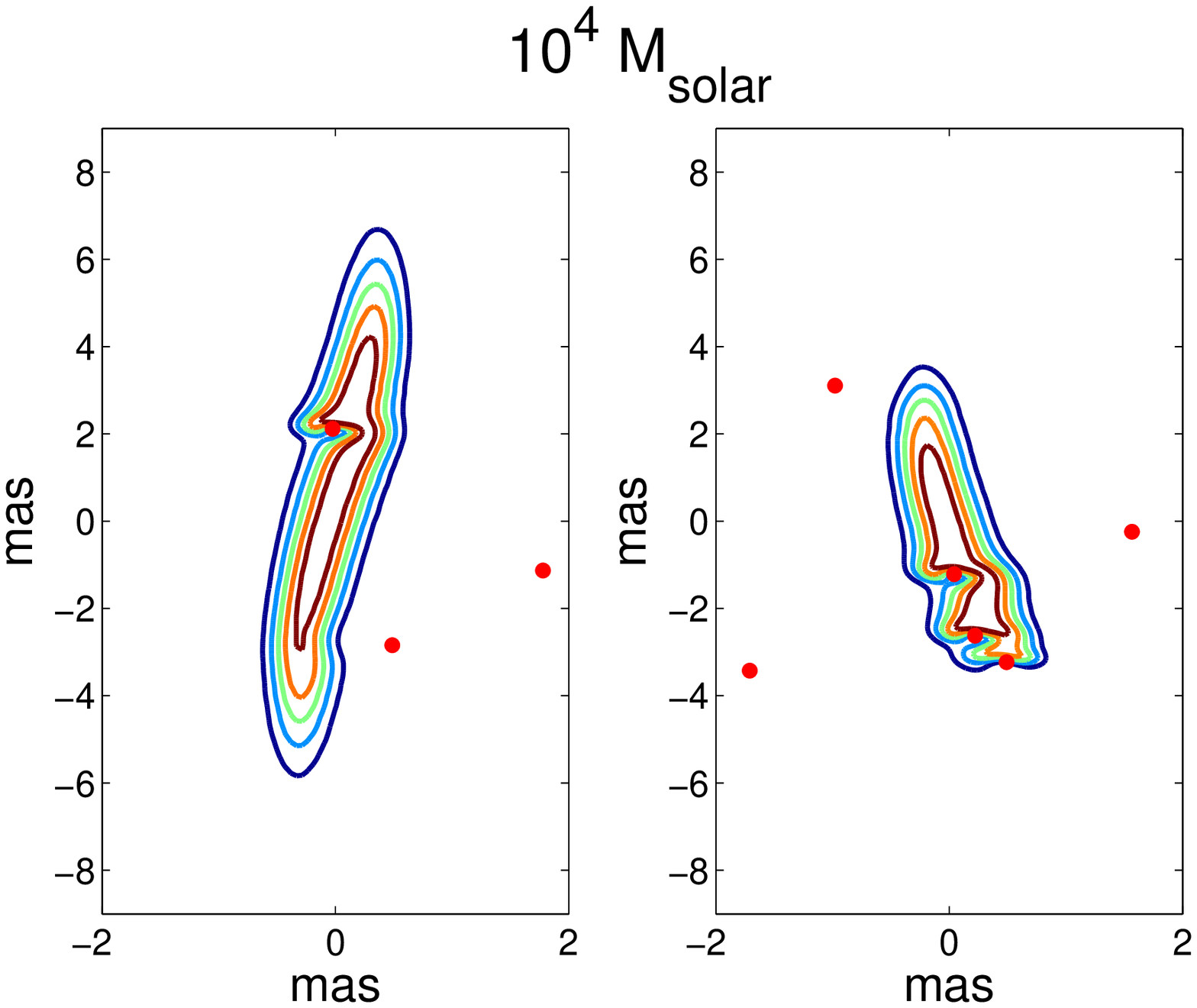}
\includegraphics[scale=0.3]{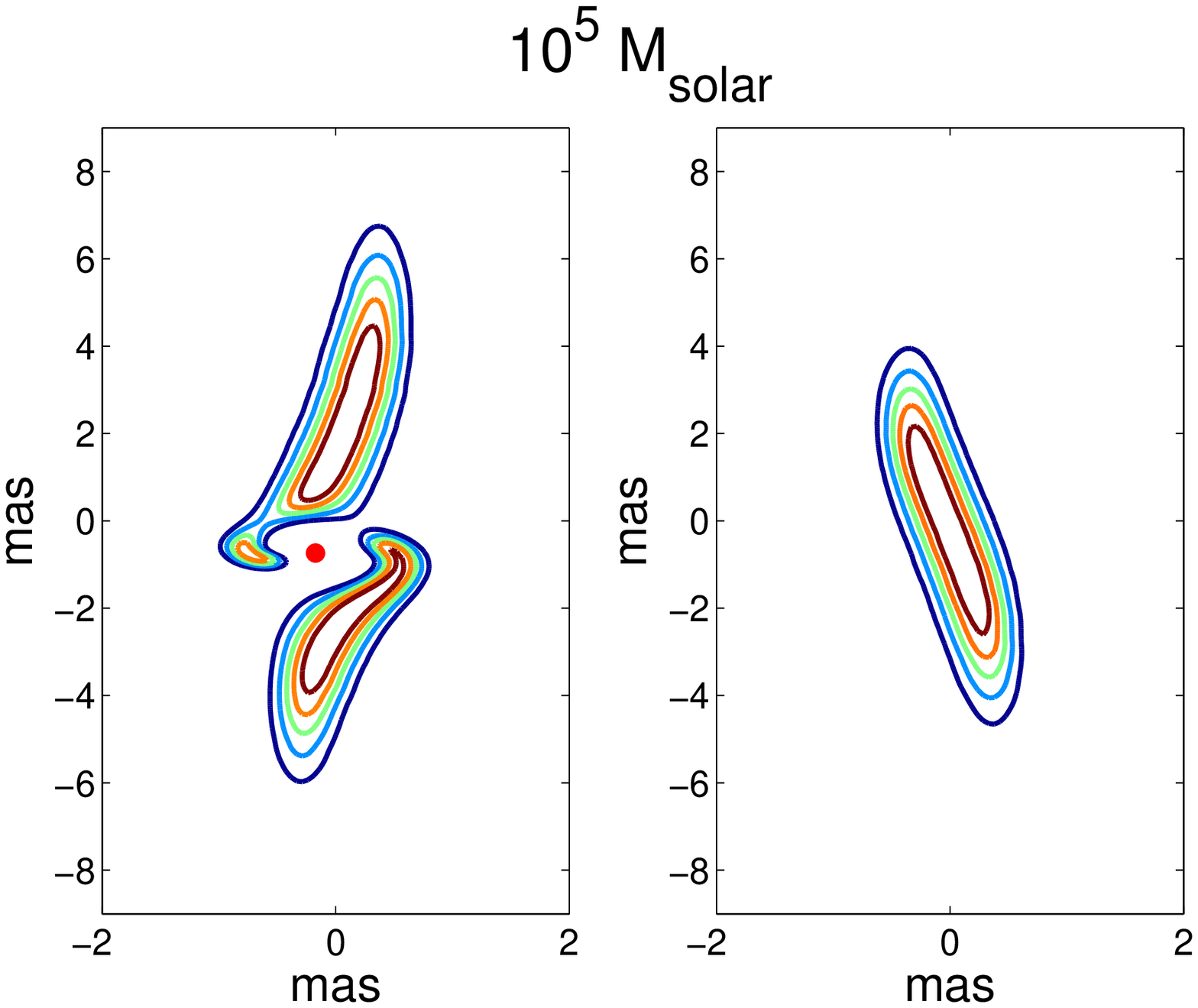}
\includegraphics[scale=0.3]{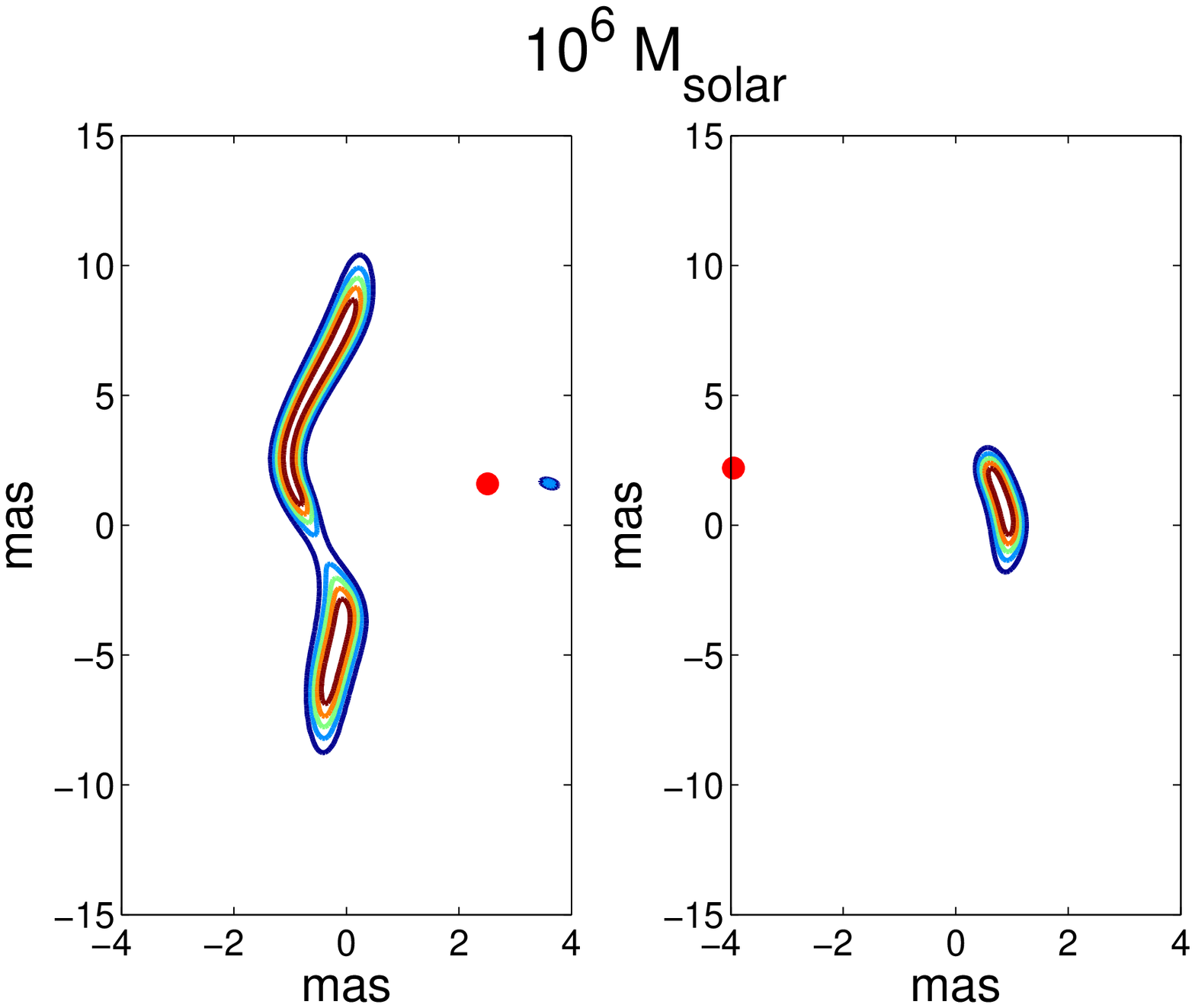}
\caption{Examples of simulated radio maps of a macrolensed quasar jet at 22 GHz (source size $10\times 2.5$ pc and resolution 0.3 milliarcseconds), subject to millilensing distortions by IMBHs with $f_\mathrm{IMBH}=0.01$ in the halo of the main lens. Each image pair represents the two macroimages from Fig.~\ref{imagemap_nosub_all}, distorted by millilensing effects from IMBHs with either $M_\mathrm{IMBH}=10^4$, $10^5$ or $10^6\ M_\odot$. The positions of the IMBHs are indicated by red dots.}
\label{imagemaps_IMBH_22GHz}
\end{figure*}

\begin{figure*}
\includegraphics[scale=0.4]{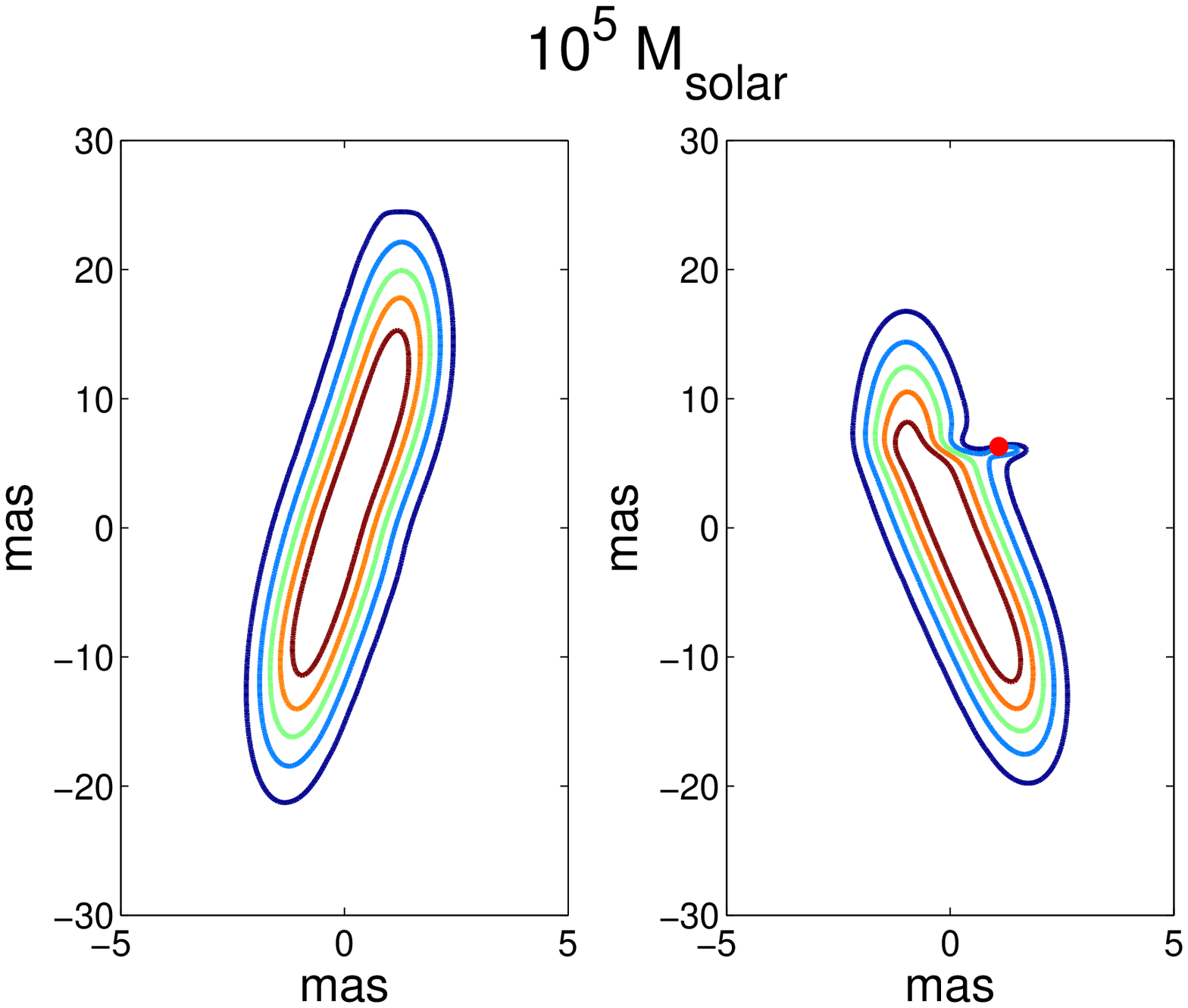}
\includegraphics[scale=0.4]{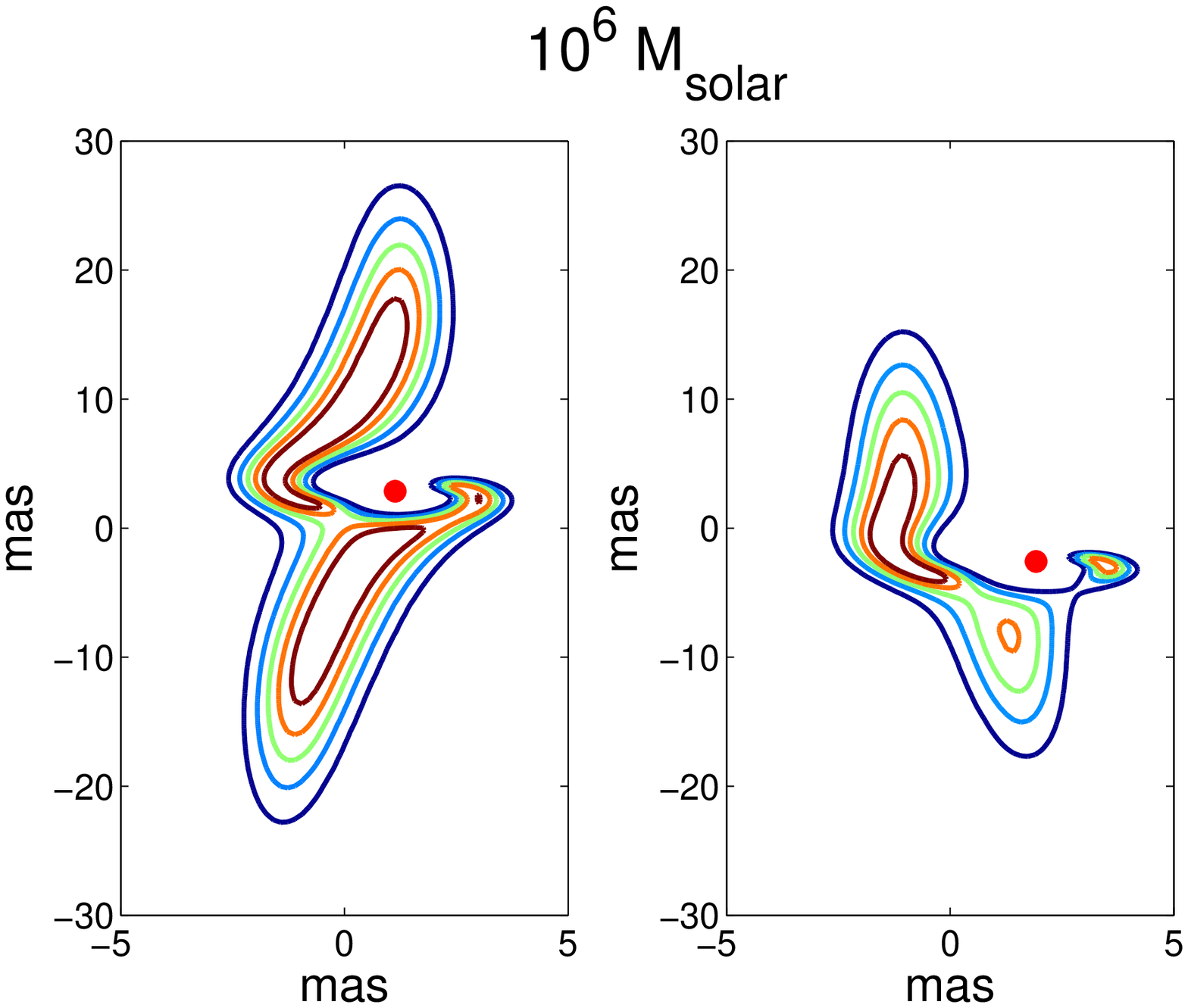}
\caption{Examples of simulated radio maps of a macrolensed quasar jet at 8.4 GHz (source size $40\times 10$ pc and resolution 0.7 milliarcseconds), subject to millilensing distortions by IMBHs with $f_\mathrm{IMBH}=0.005$ in the halo of the main lens. Each image pair represents the two macroimages from Fig.~\ref{imagemap_nosub_all}, distorted by millilensing effects from IMBHs with either $M_\mathrm{IMBH}=10^5$ or $10^6\ M_\odot$. The positions of the IMBHs are indicated by red dots.}
\label{imagemap_IMBH_8.4GHz}
\end{figure*}

In Fig.~\ref{imagemap_nosub_all}, we present our simulated images of strongly lensed quasar jets at 86, 22 and 8.4 GHz. Each image pair in the figure corresponds to the two macroimages of a single radio jet as produced by the main lens in the absence of any millilensing effects. While initially generated using a much smaller pixel scale, these images have been degraded using a Gaussian filter to match the resolution relevant for observations at these frequencies (0.05, 0.3 and 0.7 milliarcseconds respectively). The bottom row shows the corresponding isoflux contour plots, where the outermost contours correspond to $\approx 10\%$ of the peak flux in these images. All subsequent figures depict how these contours are distorted by various kinds of halo substructure within the main lens. 
Around 30 simulated lensing maps of this type are generated for each combination of $M_\mathrm{IMBH}$ and $f_\mathrm{IMBH}$ (Sect.~\ref{IMBH_simulations}), or $M_\mathrm{UCMH}$ and $f_\mathrm{UCMH}$ (Sect.~\ref{UCMH_simulations}), with randomized substructure positions for each realization, in the derivation of the detection probabilities.
 
\subsection{Detecting intermediate-mass black holes}
\label{IMBH_simulations}
In Fig.~\ref{imagemap_IMBH_86GHz} we present examples of the simulated macroimages in the case where a fraction $f_\mathrm{IMBH}=0.02$ of the dark halo of the main lens is in the form of intermediate-mass black holes with mass $M_\mathrm{IMBH}=10^{3} - 10^{6}\ M_\odot$. In this case, 86 GHz observations (ALMA + global array) are assumed, implying the smallest jet size (intrinsic length 2 pc) and the highest resolution (0.05 milliarcseconds) considered in this paper. Since $f_\mathrm{IMBH}$ is kept fixed, the number of IMBHs per unit area drops by a factor of $10^3$ in the lens plane when going from $M_\mathrm{IMBH}=10^3\ M_\odot$ to $10^6\ M_\odot$. However, because the more massive IMBHs also have larger Einstein radii, potentially detectable distortions are produced in all the cases plotted. Since the distortions in the two macroimages are uncorrelated, millilensing should also be straightforward to separate from intrinsic jet features (but see Sect.~\ref{discussion} for potential caveats).

The probability of seeing millilensing effects in at least one macroimage of a given two-image system depends on $M_\mathrm{IMBH}$, the angular resolution and $f_\mathrm{IMBH}$, but is deemed to be $P_\mathrm{milli}\gtrsim 50\%$ in all the simulations presented in Fig.~\ref{imagemap_IMBH_86GHz}. 

When analysing a survey of $N$ such macrolens systems, the probability $P_\mathrm{detection}$ of detecting millilensing becomes $P_\mathrm{detection}=1-(1-P_\mathrm{milli})^N$. By adopting $P_\mathrm{milli}\gtrsim 50\%$, one should therefore be able to rule out $f_\mathrm{IMBH}=0.02$ for IMBHs in the mass range $M_\mathrm{IMBH}=10^{3}$--$10^6\ M_\odot$ at the $\gtrsim 95\%$ level by surveying $N\approx 5$ systems. By further increasing the size of the survey, even lower $f_\mathrm{IMBH}$ can in principle be probed. To first order $P_\mathrm{milli}$ scales with $f_\mathrm{IMBH}$, so that $P_\mathrm{milli}\gtrsim 5\%$ if $f_\mathrm{IMBH}\sim 0.002$. Hence, to reach a detection probability of $P_\mathrm{detection}\gtrsim 68\%$ if $f_\mathrm{IMBH}=0.002$, one needs to observe $N\geq 22$ systems.

As demonstrated in Fig.~\ref{imagemaps_IMBH_22GHz}, the larger source size assumed for the 22 GHz (EVN) observations (10 pc) allows IMBHs with dark matter fractions as low as $f_\mathrm{IMBH}=0.01$ to be detected with probability $P_\mathrm{milli}\gtrsim 50\%$. The lower resolution (0.3 milliarcseconds) provided by the EVN at the same time prohibits the detection of IMBHs with mass $M_\mathrm{IMBH} \sim 10^3\ M_\odot$. By surveying $N\approx 5$ systems, one should be able to rule out $f_\mathrm{IMBH}=0.01$ for $M_\mathrm{IMBH} \sim 10^4$--$10^6\ M_\odot$ at 95\% confidence level. A detection probability of $P_\mathrm{detection}\gtrsim 68\%$ can also be reached at $f_\mathrm{IMBH}=0.001$ if one is able to observe $N\geq 22$ systems.

Similarly, the even larger jet (intrinsic length 40 pc) adopted for our simulated 8.4 GHz observations allows for stronger constraints on $f_\mathrm{IMBH}$, but the lower resolution (0.7 milliarcseconds) at the same time limits the IMBH mass range for which millilensing effects can be detected. Still the macroimage distortions produced by $10^5$--$10^6\ M_\odot$ IMBHs would be detectable with this resolution, and such effects would turn up with $P_\mathrm{milli}\gtrsim 50\%$ probability even if the IMBH halo mass fraction is as low as $f_\mathrm{IMBH}=0.005$. Fig.~\ref{imagemap_IMBH_8.4GHz} includes an example of such millilensing distortions produced by $10^5 M_\odot$ and $10^6 M_\odot$ IMBHs with $f_\mathrm{IMBH}=0.005$.

Table~\ref{IMBH_limits} summarizes the $f_\mathrm{IMBH}$ limits that observations of a single macrolensed jet (one image pair) at 86, 22 and 8.4 GHz would be able to probe (with $\geq 50\%$ detection probability). As further discussed in Sect.~\ref{size}, these limits can easily be rescaled to accommodate observations of larger number of multiply-imaged systems. The constraints resulting from a survey of $N\approx 5$ systems would produce constraints that are a factor of a few better than the \citet{Wilkinson et al.} millilensing constraints on $10^6\ M_\odot$ primordial black holes\footnote{Formally, the \citet{Wilkinson et al.} constraints apply to millilenses located anywhere along the line of sight to the radio sources in their sample (mean redshift $z\approx 1.3$), whereas ours apply only to millilenses within the main lens. The difference may be relevant in scenarios in which the IMBHs do not follow the distribution of dark matter on large scales (e.g. if they are formed through baryonic processes in the vicinity of galaxies.). Moreover, our approach could in principle produce somewhat stronger constraints if we were to consider millilenses along the entire line of sight}. There are no competitive {\it lensing} constraints on IMBH at $10^3$--$10^5 \ M_\odot$, but there are still a host of other constraints that may be applicable, in particular those related to accretion onto these objects \citep[see][ for a review]{Carr et al.}.

\begin{table}
\caption{The lowest halo mass fraction in IMBHs, $f_\mathrm{IMBH}$, that would produce detectable millilensing distortions with $P_\mathrm{milli}\gtrsim 50\%$ probability in a single macroimage pair.}
\begin{tabular}{@{}llll@{}}
\hline
Frequency (GHz) & Mass ($M_\odot$) & Source size (pc) & min $f_\mathrm{IMBH}$ \\
\hline
86 & $10^3-10^6$  & $2 \times 0.5$& $2\times 10^{-2}$\\
22 & $10^4-10^6$  & $10 \times 2.5$& $1\times 10^{-2}$\\
8.4 & $10^5-10^6$  & $40 \times 10$& $5\times 10^{-3}$\\
\hline
\label{IMBH_limits}
\end{tabular}
\end{table}

\begin{table}
\caption{The lowest halo mass fraction in UCMHs, $f_\mathrm{UCMH}$, that would produce detectable millilensing distortions with $P_\mathrm{milli}\gtrsim 10\%$ probability in a single macroimage pair.}
\begin{tabular}{@{}llll@{}}
\hline
Frequency (GHz) & Mass ($M_\odot$) & Source size (pc) & min $f_\mathrm{UCMH}$ \\
\hline
86 & $10^6-10^8$  & $2 \times 0.5$& $2\times 10^{-1}$\\
22 & $10^7-10^8$  & $10 \times 2.5$& $1\times 10^{-1}$\\
8.4 & $10^8$  & $40 \times 10$& $5\times 10^{-2}$\\
\hline
\label{UCMH_limits}
\end{tabular}
\end{table}

\begin{figure*}
\includegraphics[scale=0.4]{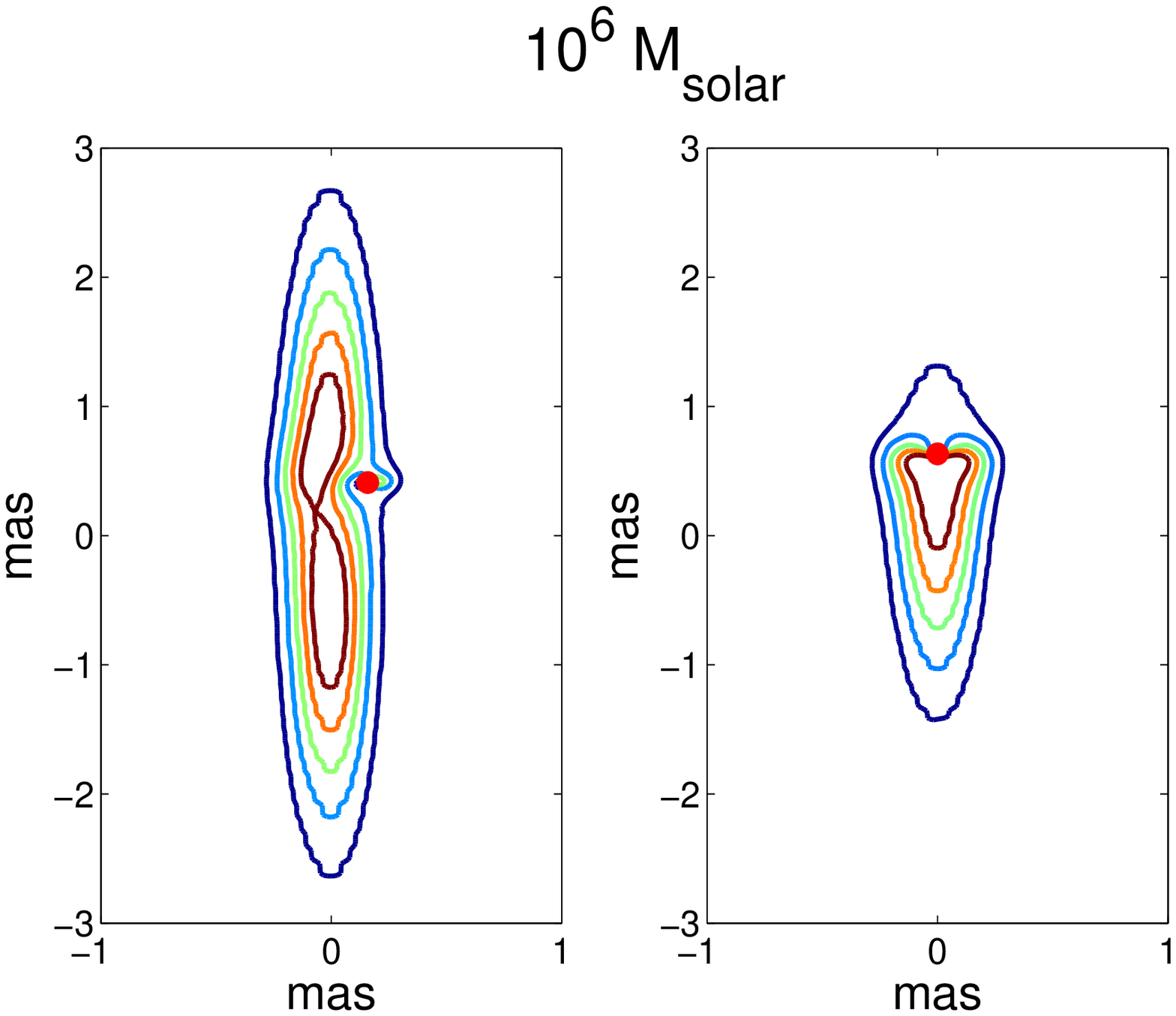}
\includegraphics[scale=0.4]{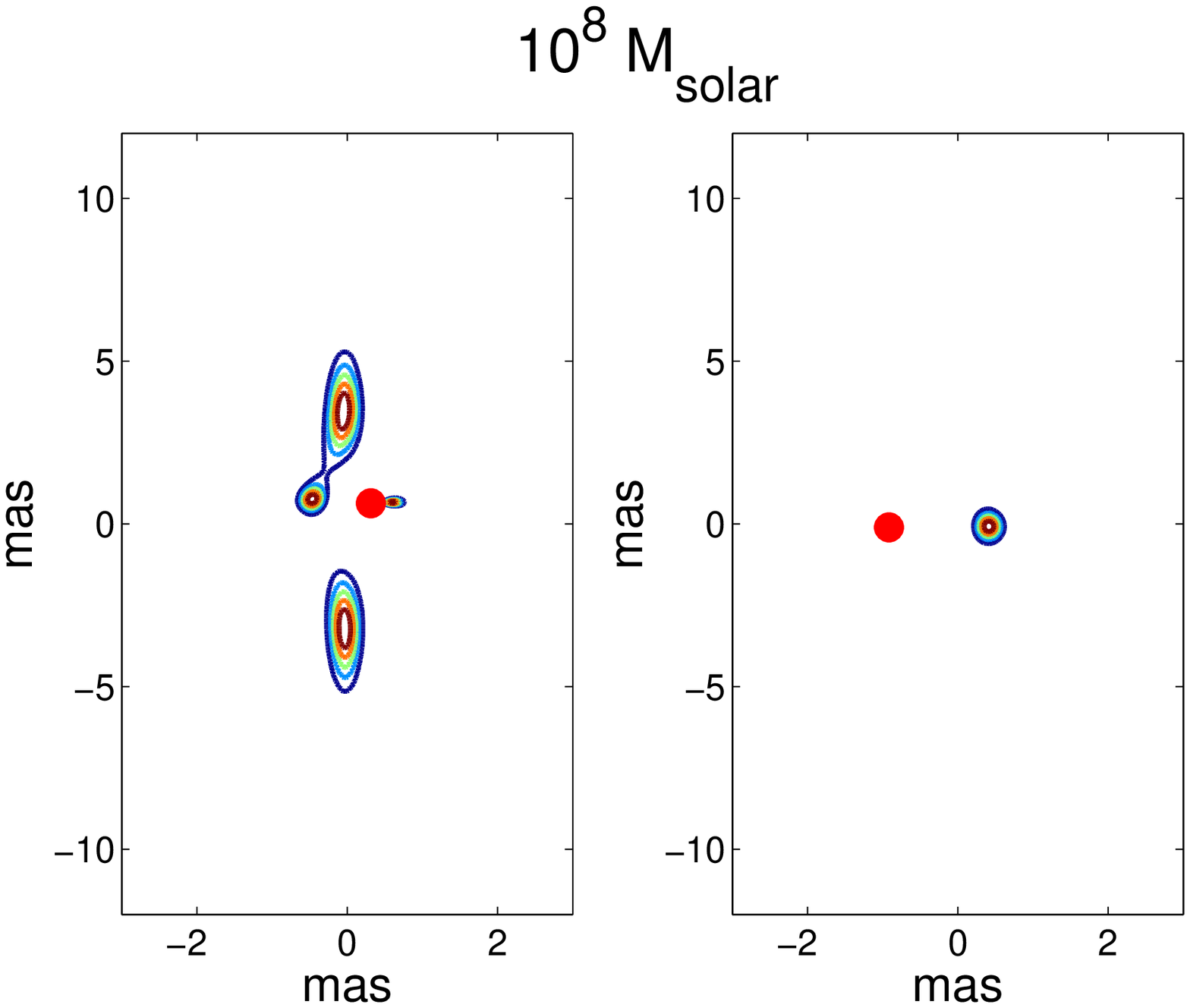}
\caption{Examples of simulated radio maps of a macrolensed quasar jet at 86 GHz (source size $2\times 0.5$ pc and resolution 0.05 milliarcseconds), by UCMHs with $M_\mathrm{UCMH}=10^6$ and $10^8 \ M_\odot$  in the halo of the main lens. The probabilities of detecting such effects are, however, negligibly small unless the UCMH dark halo fraction is $f_\mathrm{UCMH}\sim 0.2$.}
\label{imagemap_UCMH_86GHz}
\end{figure*}

\subsection{Detecting ultracompact minihalos}
\label{UCMH_simulations}
When compared to IMBHs of the same mass, UCMHs have much smaller Einstein radii and are far more difficult to detect through 
millilensing effects. Our simulations show, that while $M_\mathrm{UCMH}\sim 10^6$--$10^8\ M_\odot$ UCMHs may in principle be detectable through small-scale macroimage distortions, the probability of observing this effect is exceedingly small unless the UCMH dark matter fraction $f_\mathrm{UCMH}$ is very high. 

In Fig.~\ref{imagemap_UCMH_86GHz}, we show examples of the millilensing distortions that $10^6$ and $10^8\ M_\odot$ UCMHs would produce in the case of 86 GHz observations (ALMA + the global array). However, the probability of seeing effects of this type in a given macroimage pair is only $P_\mathrm{milli}\approx 10\%$, even for a UCMH dark halo fraction as high as $f_\mathrm{UCMH}=0.2$. The detection prospects become somewhat better at 22 and 8.4 GHz ($f_\mathrm{UCMH}\gtrsim 0.05$-0.1 at $P_\mathrm{milli}\approx 10\%$) due to the larger source adopted sizes at these frequencies, but only for $10^7$--$10^8\ M_\odot$ objects (see Table~\ref{UCMH_limits}). 

By probing $N\approx 11$ (28) macroimage pairs, the detection probability can be pushed to $P_\mathrm{detection}\approx 68\%$ ($95\%$) at these $f_\mathrm{UCMH}$ limits. In order to probe UCMH dark halo fractions significantly below $f_\mathrm{UCMH}\sim 0.1$, hundreds of images would therefore need to be observed. While there are no competitive lensing constraints at $\sim 10^6 \ M_\odot$, it is possible that the \citet{Wilkinson et al.} observations of 300 $z\sim 1$ radio sources (not macrolensed) at $\sim 1$ milliarcsecond resolution would be able to do better for $\sim 10^8 \ M_\odot$ UCMHs than the predicted limits we give in this paper.

\subsection{Detecting standard CDM subhalos}
\begin{figure*}
\includegraphics[scale=0.4]{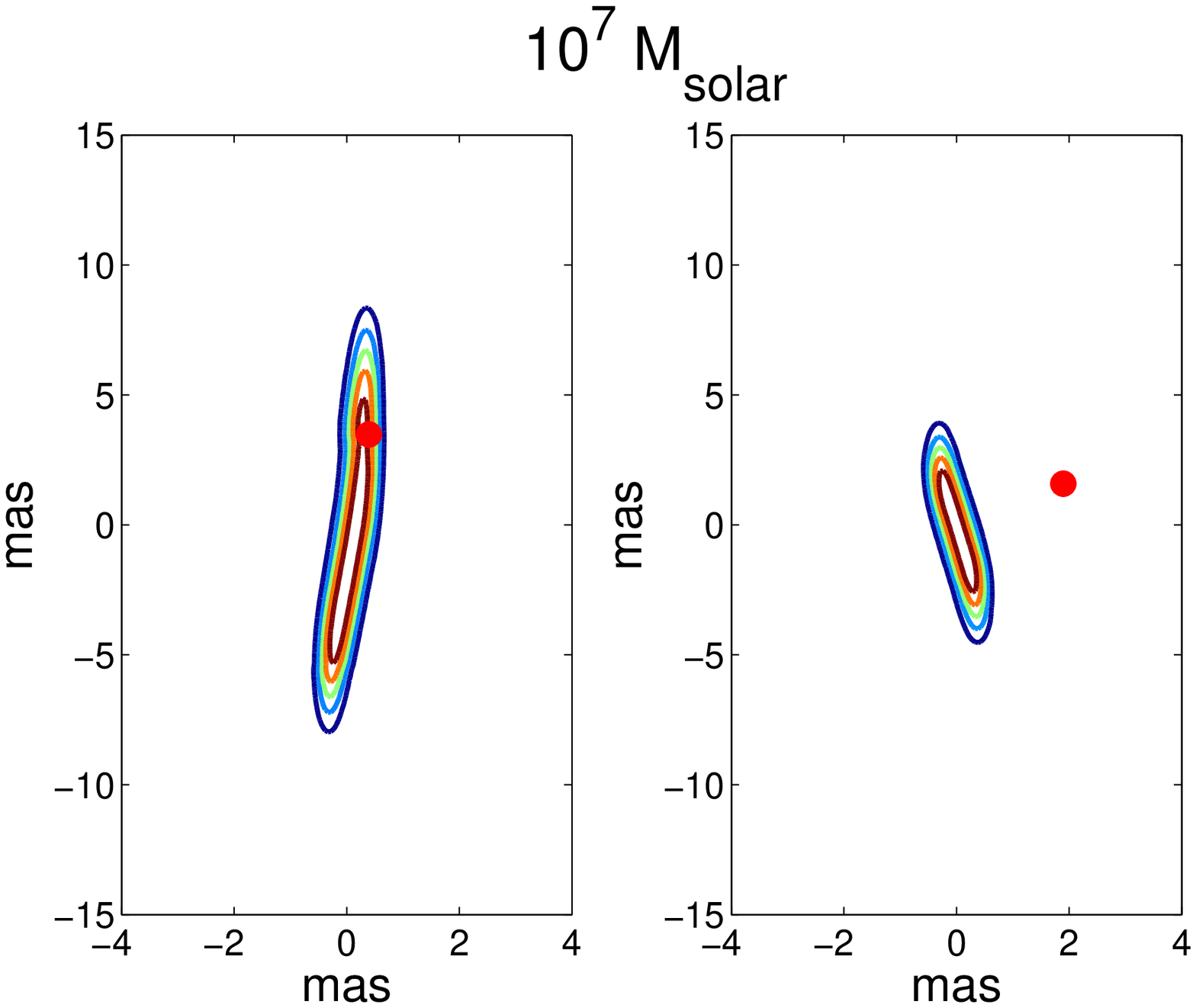}
\includegraphics[scale=0.4]{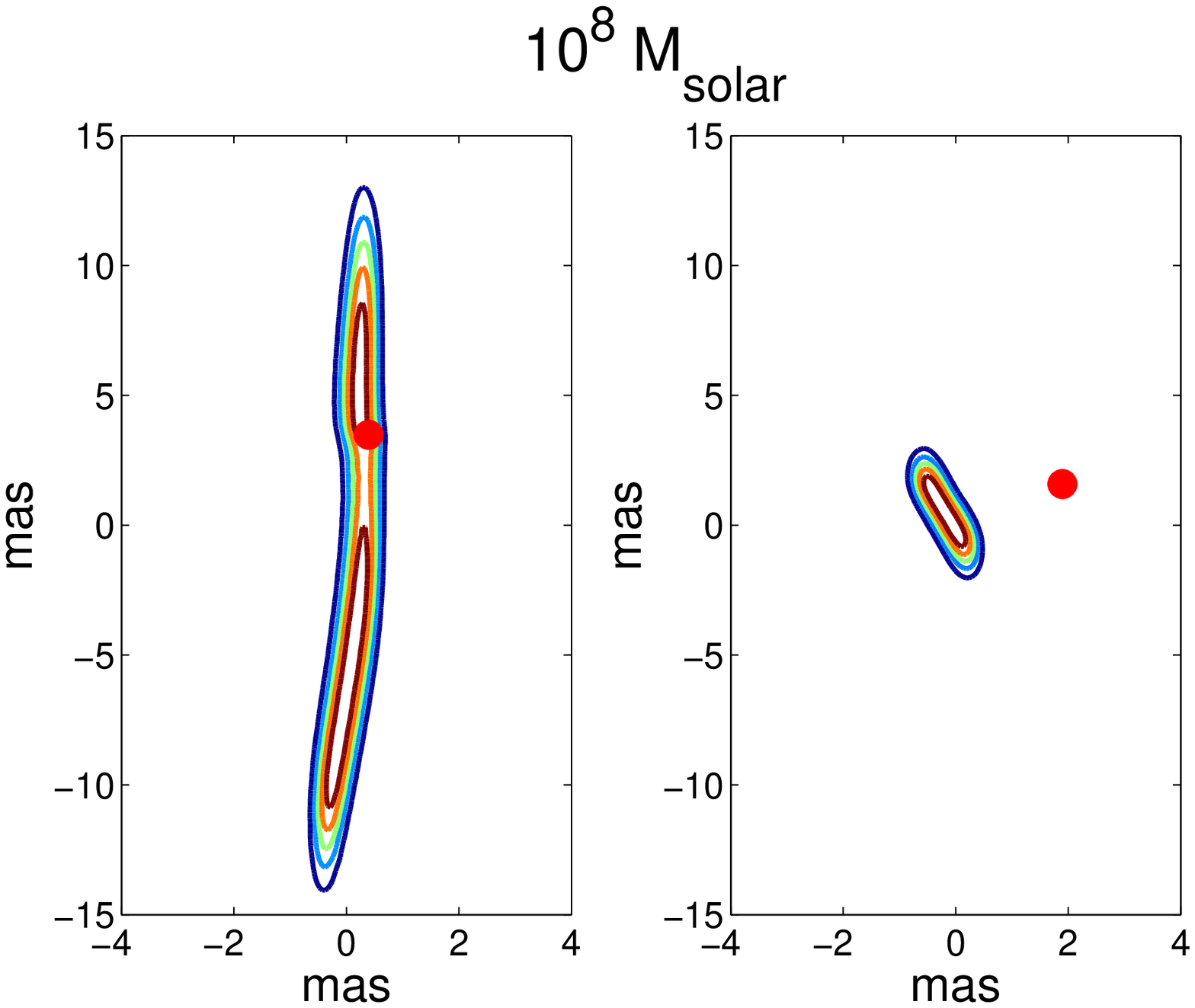}
\caption{Examples of simulated radio maps of a macrolensed quasar jet at 22 GHz (source size $10\times 2.5$ pc and resolution 0.3 milliarcseconds), subject to millilensing distortions by NFW subhalos with masses of either $\sim 10^7$ or $10^8\ M_\odot$. The red dots mark the positions of the centres of these subhalos. As the left pair of images shows, $\sim 10^7\ M_\odot$ NFWs produce very mild distortions only, whereas NFW subhalos of mass $\sim 10^8\ M_\odot$ may produce more significant distortions if they are placed sufficiently close to a macroimage. The probability for such superpositions to occur is, however, very small.}
\label{imagemap_NFW}
\end{figure*}

As expected, standard CDM subhalos (assumed to have NFW density profiles) are not detectable using the observational scheme considered in this paper. In Fig.~\ref{imagemap_NFW}, we show examples of the $\sim 10^7$ and $\sim 10^8\ M_\odot$ subhalos close to the macroimages in our simulations at 22 GHz, assuming the \citet{Bullock et al.} relation between mass and concentration parameter. In the standard case of $f_\mathrm{sub}\approx 0.002$, only $M\lesssim 10^7 M_\odot$ subhalos are sufficiently numerous to have a decent probability of showing up in the vicinity of the macroimages, and even though such objects may affect the overall magnification and the curvature of the jet (see left panel of Fig.~\ref{imagemap_NFW}), the associated small-scale distortion is too small to be resolved. NFW subhalos at $\sim 10^8\ M_\odot$ may in principle give rise to the detectable distortions (right panel of Fig.~\ref{imagemap_NFW}), but the probability of attaining the required aligment between macroimage and subhalo is negligibly small. We estimate that the probability of detecting small-scale distortions due to $10^8 M_\odot$ NFW subhalos in a single macroimage pair is no more than $P_\mathrm{milli}\approx 3\times10^{-4}$ in this case. At $10^9 M_\odot$, the probability is even lower ($P_\mathrm{milli}\approx 4\times10^{-5}$). 

Recent results by \citet{Vegetti et al. b, Vegetti et al. c} hint at a higher surface mass density contribution ($f_\mathrm{sub}\approx 0.03$) and a flatter subhalo mass function ($\alpha=1.1$ in Eq. (\ref{subhalo mass function})) than predicted by current CDM simulations, but even if we adopt these values, the probability for detection remains too low ($\approx 5\times 10^{-4}$ for NFWs of mass $10^8$--$10^9\ M_\odot$) to make this search strategy attractive.

The problem is one of source size -- the macrolensed jets we consider cover an area in the lens plane that is several orders of magnitude too small to intersect such massive subhalos. The intrinsic source size would need to have an area $\sim 10^3$ times greater than the largest jets we consider ($40\times 10$ pc at 8.4 GHz) to push the detection probabilities into the interesting range $P_\mathrm{milli}\gtrsim 10\%$. This essentially requires a completely different kind of source, like the dusty sub-mm galaxies considered by \citet{Inoue & Chiba b}.

These result are admittedly sensitive to the concentration parameters adopted for the NFW subhalos. In the examples above, we have used the $c(M_\mathrm{vir})$ relation from \citet{Bullock et al.}, which for objects in the relevant mass range ($10^7$--$10^8\ M_\odot$) results in concentration parameters a factor of $\approx 2$ higher than the ones predicted by the \citet{Maccio et al. a} relation. If we instead adopt the \citet{Maccio et al. a} $c(M_\mathrm{vir})$ scaling, the detection threshold shifts upward by an order of magnitude in mass, so that image distortions predicted for $10^7$ and $10^8\ M_\odot$ NFW in the Bullock et al. case (Fig.~\ref{imagemap_NFW}) instead are produced at masses of $\sim 10^8$ and $\sim 10^9\ M_\odot$. 

\section{Discussion}
\label{discussion}

\subsection{Temporal effects}
In previous sections, we have argued that millilensing-induced distortions of quasar jets may be distinguished from morphological features intrinsic to these sources, since the latter would be reproduced in all macroimages whereas millilensing should affect each image independently. However, this argument comes with a caveat. The time delay between the images in quasar-galaxy lenses can be up to a year \citep[for a compilation of time delays, see][]{Oguri}, which means that intrinsic, transient features in the jet may, at any given time, be visible in just one of the images and be mistaken for millilensing effects. This is for instance likely to be the case in superluminal radio jets, where blobs are seen to move $\sim 1$ milliarcseconds yr$^{-1}$ along the jet \citep[e.g.][]{Jorstad et al.}. For macrolensed jets that show signs of millilensing distortions, it may therefore become necessary to obtain data at two or more epochs. Since halo substructures give rise to millilensing magnification pattern that will appear stationary over decades \citep{Metcalf & Madau}, any distortions that seem to move along the jet are bound to be intrinsic to the source. Small-scale features that are not duplicated in the other macroimages and appear with a fixed angular position (as, for instance, measured from the base of the jet) over the course of more than a year is on the other hand likely caused by millilensing.  
\begin{table}
\caption{Impact parameter $R_\mathrm{eff}$ within which a subhalo of a given type will produce detectable macroimage distortions}
\begin{tabular}{@{}lllll@{}}
\hline
Type & Resolution    & Mass        & $\mu$ & $R_\mathrm{eff}$ \\
     & (milliarcsec) & ($M_\odot$) &       & (pc)\\
\hline
IMBH & 0.05 (86 GHz) & $10^3$ & 3  & 1\\ 
     & 							 & $10^4$ &    & 3\\
     & 							 & $10^5$ &    & 7\\
     & 							 & $10^6$ &    & 20\\   
     &               & $10^3$ & 10 & 2\\ 
     & 							 & $10^4$ &    & 6\\
     & 							 & $10^5$ &    & 20\\
     & 							 & $10^6$ &    & 50\\   
     &               & $10^3$ & 30 & 2\\ 
     & 							 & $10^4$ &    & 10\\
     & 							 & $10^5$ &    & 40\\
     & 							 & $10^6$ &    & 80\\   
     & 0.3 (22 GHz)	 & $10^4$ & 3  & 2\\   
     & 							 & $10^5$ &    & 5\\
     & 							 & $10^6$ &    & 20\\   
     &    	 				 & $10^4$ & 10 & 2\\   
     & 							 & $10^5$ &    & 7\\
     & 							 & $10^6$ &    & 30\\   
     &    	 				 & $10^4$ & 30 & 2\\   
     & 							 & $10^5$ &    & 8\\
     & 							 & $10^6$ &    & 40\\   
     & 0.7 (8.4 GHz) & $10^5$ & 3  &  3\\   
     & 							 & $10^6$ &    & 10\\
     &    	 				 & $10^5$ & 10 &  4\\   
     & 							 & $10^6$ &    & 20\\
     &    	 				 & $10^5$ & 30 &  8\\   
     & 							 & $10^6$ &    & 40\\
UCMH & 0.05 (86 GHz) & $10^6$ & 3  &  2\\ 
     & 							 & $10^7$ &    &  6\\
     & 							 & $10^8$ &    &  20\\
     &               & $10^6$ & 10 &  3\\ 
     & 							 & $10^7$ &    &  20\\
     & 							 & $10^8$ &    &  60\\
     &               & $10^6$ & 30 &  10\\ 
     & 							 & $10^7$ &    &  30\\
     & 							 & $10^8$ &    &  100\\
     & 0.3 (22 GHz)	 & $10^7$ & 3  &  2\\   
     & 							 & $10^8$ &    &  10\\
     &    	 				 & $10^7$ & 10 &  4\\   
     & 							 & $10^8$ &    &  30\\
     &    	 				 & $10^7$ & 30 &  5\\   
     & 							 & $10^8$ &    &  60\\
     & 0.7 (8.4 GHz) & $10^8$ & 3  &  10\\   
     &    	 				 & $10^8$ & 10 &  20\\   
     &    	 				 & $10^8$ & 30 &  60\\   
\hline
\label{Reff}
\end{tabular}
\end{table}

\subsection{Source size sensitivity}
\label{size}
\begin{figure}
\includegraphics[scale=0.55]{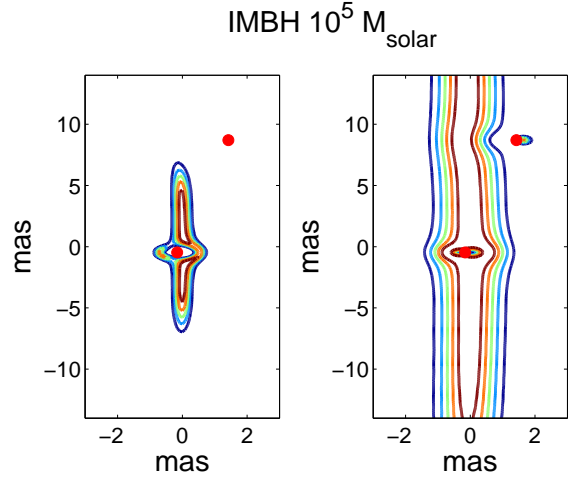}
\caption{Illustration of how source size affects the probability for detecting dark halo substructure. The two frames depict a single macroimage (out of a two-image pair) with fixed macrolensing magnification ($\mu=10$) but with different intrinsic source size: $10 \times 2.5$ pc (left) and $40 \times 10$ pc (right). The smaller version (left) corresponds to the source size adopted for our 22 GHz simulations (see Sect.~\ref{source size}). The red dots mark the positions of two $10^5 \ M_\odot$ IMBHs (identical positions in both frames). In the small-source case (left), only one of these IMBHs produce detectable millilensing effects, whereas both can be detected in the large-source case (right) due to better macroimage coverage of the lens plane. A resolution of 0.3 milliarcsec has been adopted in both cases (as considered suitable for 22 GHz observations).
\label{size_example}}
\end{figure}

For a fixed substructure type and telescope beam size, the prospects of detecting millilensing effects depend on the adopted source dimensions. This is exemplified in Fig.~\ref{size_example}, where the probability of substructures is seen to increase with source area -- whereas only one IMBH is detectable in small-source case (left), two IMBHs are detectable for the larger source (right). The source sizes adopted in Sect.~\ref{source size} are uncertain by a factor of a few, and it may be convenient to be able to generalize our results to match other source dimensions. For a given substructure type and mass, any detection limit $\min(f_\mathrm{sub, 1})$ (as listed in Table ~\ref{IMBH_limits} and ~\ref{UCMH_limits}) derived for an intrinsically elliptical source with area $A_1$ can be rescaled to some other source area  $A_2$ using:
\begin{equation}
\min f_\mathrm{sub, 2} \approx \frac{A_1 + C_1 R_\mathrm{eff}}{A_2 + C_2 R_\mathrm{eff}} \min f_\mathrm{sub, 1}
\label{rescaling_formula}
\end{equation}
Here, $\min f_\mathrm{sub, 2}$ is the rescaled detection limit relevant for source area $A_2$, whereas the $C$ parameters represent the circumferences of the macorimages one wants to rescale from ($C_1$) and to ($C_2$). The impact parameter  $R_\mathrm{eff}$ measures the  projected distance from the subhalo centre within which detectable macroimage distortions will be produced. This impact parameter, which depends on both subhalo mass and type, is typically larger than the subhalo Einstein radius, since substantial deflection can occur even outside the latter. The $R_\mathrm{eff}$ values relevant for $10^3$--$10^6\ M_\odot$ IMBHs and $10^6$--$10^8\ M_\odot$ UCMHs are listed in Table~\ref{Reff} for the resolutions adopted at 8.4, 22 and 86 GHz. Since these $R_\mathrm{eff}$ values also depend on the magnification of the macroimage, $R_\mathrm{eff}$ values are presented for $\mu=3$, 10 (our default value) and 30. 

This rescaling scheme, which assumes that the source size and $R_\mathrm{eff}$ are independent, is admittedly an approximation and reliable only to within a factor of a few. Secondary images due to substructure lensing may for instance be easier to detect for a compact rather than an extended source due to flux ratio issues. An effect of the latter type is evident in Table~\ref{Reff}, where both $10^6$ IMBHs and $10^8$ UCMH are seen to have larger $R_\mathrm{eff}$ at 86 GHz (smallest source) than at 22 (intermediate source) or 8.4 GHz (largest source).

\begin{figure*}
\includegraphics[scale=0.55]{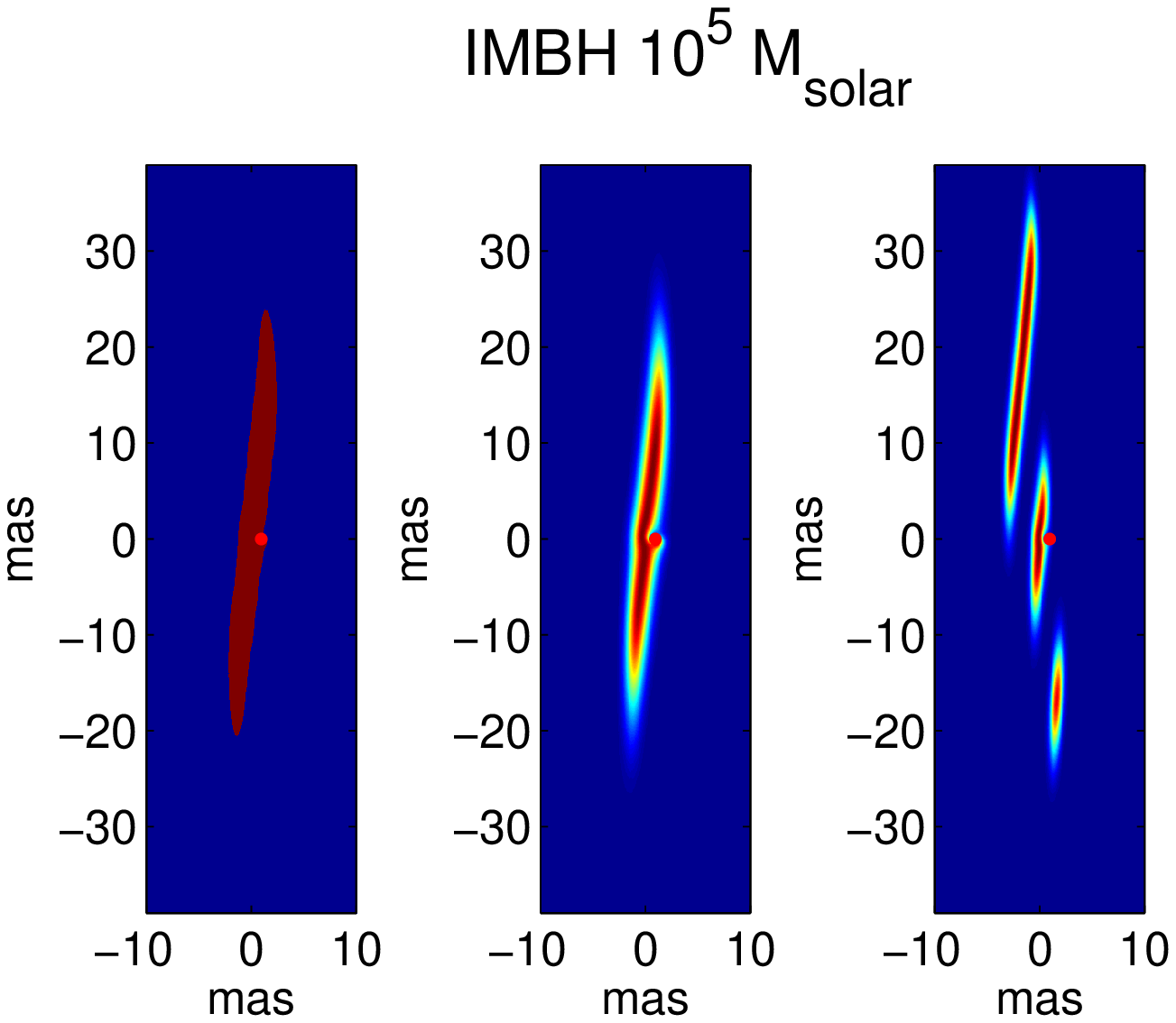}
\includegraphics[scale=0.55]{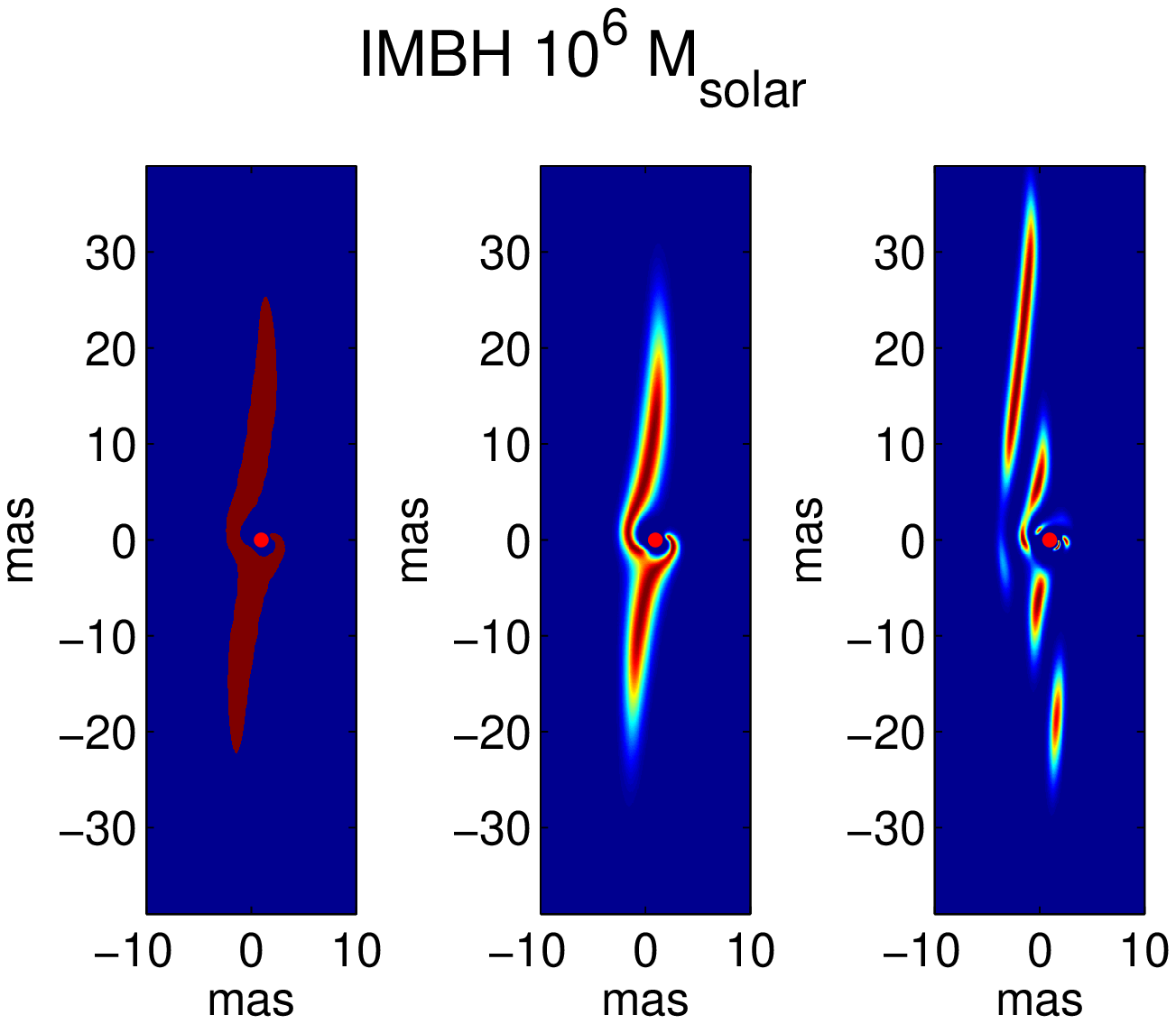}
\caption{Examples of how source morphology and surface brightness distribution affect the detectability of millilensing from IMBHs of mass $10^5\ M_\odot$ and $10^6\ M_\odot$ against a simulated radio map of a strongly lensed quasar jets at 8.4 GHz. For each IMBH mass, the leftmost frame contains a source with constant surface brightness, the middle one the Gaussian profile used throughout the rest of the paper, and the rightmost frame a source consisting of a sequence of Gaussian ``blobs" of different sizes. Contour representations are not used in this plot, since this becomes confusing in the case of a flat surface brightness profile. In general, complicated source morphologies (rightmost frames) do not significantly compromise the detectability of millilensing effects, but sources with shallow (or even constant) surface brightness profiles may render certain forms of substructure lenses undetectable (as seen in the case where $10^5\ M_\odot$ IMBH are superposed on a constant surface brightness source).}
\label{sfb}
\end{figure*}

\subsection{The surface brightness profile}
\label{sbp}
In previous sections, we have assumed the source to be an intrinsically straight jet with a surface brightness distribution described by a 2-dimensional Gaussian. While the intrinsic source morphology and surface brightness profile is less important than the overall source area when assessing lensing probabilities, there are certain situations where they do matter. Since gravitational lensing conserves surface brightness, halo substructure can only produce detectable image distortions if it happens to affect a region of the macroimage where there is a non-negligble surface brightness gradient. In the extreme case of a source with constant surface brightness, halo substructure will not produce {\it any} detectable features unless its lensing effects extends beyond the macroimage boundary. This is exemplified in Fig.~\ref{sfb}, where $10^5\ M_\odot$ and $10^6\ M_\odot$ IMBHs (red dots) are superposed on macroimages of (from left to right, for each IMBH mass) an elliptical source with constant surface brightness, an elliptical source with Gaussian surface brightness profile and a patchy jet with Gaussian ``blobs" of increasing size when moving from the lower-right to upper-left corner. 

In this example, a $10^5\ M_\odot$ IMBH becomes undetectable in the case of a constant surface brightness source, but can be spotted as a mild distortion against the Gaussian source. An IMBH of this mass redistributes surface brightness within an area that is much smaller than that of the source. Hence, if the source surface brightness is constant, no detectable effects are produced. Even though placed in the exact same position, the lensing produced by a $10^6\ M_\odot$ IMBH on the other hand extends sufficiently far out to distort the rim of the macroimage and can therefore be detected regardless of the source profile. In fact, the only $f_\mathrm{IMBH}$ entry in Table ~\ref{IMBH_limits} that would change in any dramatic way when going from a Gaussian to a constant surface brightness source corresponds to the $10^5\ M_\odot$ case depicted in Fig.~\ref{sfb} (i.e. source size and resolution corresponding to 8.4 GHz). In this case, constant surface brightness source would effectively prevent any useful $f_\mathrm{IMBH}$ constraints, whereas the changes are modest in all other cases. Since UCMH lenses produce more long-range effects then IMBHs, the $f_\mathrm{UCMH}$ estimates in Table ~\ref{UCMH_limits} are even less affected by the source surface brightness profile.

Fig.~\ref{sfb} also provides an example of a more patchy jet morphology. This jet has the same source area as the other source cases, and consequently extends further in the vertical direction due to the empty regions between the ``blobs". Both $10^5\ M_\odot$ and  $10^6\ M_\odot$ IMBHs are in principle detectable against the source in this example, although the distortion produced in the former case becomes very modest since the IMBH happens to be projected on the outskirts of one of the blobs. In general, having a complicated jet morphology does not significantly compromise the detectability of millilensing effects. Instead, a morphology of this type could even boost the detection prospects in cases where the substructure $R_\mathrm{eff}$ (see Sect.~\ref{size}) is larger than empty regions in the macroimage (as in the $10^6\ M_\odot$ IMBH case in Fig.~\ref{sfb}), since the effective source area becomes larger in this situation.

\subsection{The nature of the substructures}
The detection of milliarcsecond or submilliarcsecond-scale image distortions would prove the existence of substructures within the macrolens, and also allow constraints on their surface number densities (as a function of substructure mass and type) to be set. However, the exact nature of a single millilens may still be very difficult to determine, since a low-mass, high-density substructure can produce a distortion very similar to that of a high-mass, low-density object. While \citet{Inoue & Chiba c} have demonstrated that the distortions induced in extended images (like the ones we model here) contain some information about the density profiles of the lenses, the finite resolution and sensitivity of actual observations could still allow for considerable degeneracies in cases where neither the masses nor the density profiles of the millilenses are known. IMBHs and UCMHs can for instance produce very similar lensing distortions in our simulations (although at different masses -- a UCMH typically needs to be a factor of $\sim 10^3$ more massive than an IMBH to reproduce a given feature). While it is possible that a combined consideration of small-scale distortions (e.g. the bending of a macrolensed jets), astrometric perturbations (the positional shift of a macroimage produced by the presence of substructures) and macroimage flux ratios could provide some constraints, this is beyond the scope of the present paper.

\section{Summary}
\label{summary}
Using simulations of strongly lensed quasar jets, we argue that very dense forms of halo substructure (intermediate-mass black holes and ultracompact minihalos) within the main lens may reveal itself through small-scale morphological distortions in the macroimages. Such distortions can be distinguished from intrinsic source features by obtaining data at multiple epochs. By mapping a handful of macrolensed jet systems at submilliarcsecond resolution, we argue that $\sim 10^3$--$10^6\ M_\odot$ intermediate-mass black holes can be detected or ruled out if they contribute a surface mass fraction of $f_\mathrm{IMBH}\gtrsim 0.01$ (depending on the  VLBI array and frequency used) to the dark matter of the main lens at the macroimage positions. Ultracompact minihalos in the $\sim 10^6$--$10^8\ M_\odot$ mass range may similarily produce detectable small-scale effects if such objects comprise  $f_\mathrm{UCMH}\gtrsim 0.1$ of the dark matter. While standard CDM subhalos at masses of $\gtrsim 10^8\ M_\odot$ can in principle also produce milliarcsecond-scale distortions, provided that such objects are projected sufficiently close to the macroimages, the probability of this is too small ($P_\mathrm{milli}\sim 10^{-4}$) for sources of the type we consider.

\section{Acknowledgements}
E.Z. acknowledges funding from the Swedish Research Council and the Swedish National Space Board. P.S. is supported by the Lorne Trottier Chair in Astrophysics, and institute for Particle Physics Theory Fellowship and a Banting Fellowship, administered by the Natural Science and Engineering Research Council of Canada.

\end{document}